\documentclass [12pt]{article}
\usepackage{amsmath,amssymb,cite}

\usepackage{tabularx}
\usepackage[english]{babel}
\usepackage[dvips]{graphicx}
\usepackage{indentfirst}
\usepackage{epsfig}
\begin{document}

\title{Notes on noncommutative supersymmetric  gauge theory on the fuzzy supersphere
}

\author{Badis Ydri\footnote{Email: ydri@physik.hu-berlin.de. The Humboldt Universitat Zu Berlin preprint number is HU-EP-07/35. 
}\\
Institut fur Physik, Humboldt-universitat zu Berlin\\
Newtonstr.15, D-12489 Berlin-Germany.}

\maketitle

\begin{abstract}
In these notes we review ~Klim\v{c}\'{\i}k's construction of noncommutative gauge theory on the fuzzy supersphere. This theory has an exact SUSY gauge symmetry with a finite number of degrees of freedom and thus {\it in principle} it is amenable to the methods of matrix models and  Monte Carlo numerical simulations. We also write down in this article a novel fuzzy supersymmetric scalar action on the fuzzy supersphere.
\end{abstract}
The differential calculus on the fuzzy sphere is $3-$dimensional and as a consequence a spin $1$ vector field $\vec{C}$ is intrinsically $3-$dimensional. Each component $C_i$, $i=1,2,3$, is an element of some matrix algebra $Mat_{N}$. Thus $U(1)$ symmetry will be implemented by $U(N)$ transformations. On the fuzzy sphere $S^2_N$ it is not possible to split the vector field $\vec{C}$ in a gauge-covariant fashion into a tangent $2$-dimensional gauge field and a normal scalar fluctuation. Thus in order to reduce the number of independent components from $3$ to $2$ we impose the  gauge-covariant condition
\begin{eqnarray}
\frac{1}{2}(x_iC_i+C_ix_i)+\frac{C_i^2}{\sqrt{N^2-1}}=0.\label{act0}
\end{eqnarray}
$x_i={L_i}/{\sqrt{L_i^2}}$ ( where $L_i$ are the  generators of $SU(2)$ in the irreducible representation $\frac{N-1}{2}$ of the group ) are the matrix coordinates on fuzzy $S^2_N$. The  action on the fuzzy sphere $S^2_N$ is given by

\begin{eqnarray}
S_N[C]&=&\frac{1}{4Ng^2}TrF_{ij}^2-\frac{1}{2Ng^2}{\epsilon}_{ijk}Tr_L\left[\frac{1}{2}F_{ij}C_k-\frac{i}{6}[C_i,C_j]C_k\right].\label{act1}
\end{eqnarray}
$F_{ij}$ is the curvature on the fuzzy sphere, viz $F_{ij}=i[Y_i,Y_j]+{\epsilon}_{ijk}Y_k$ where the covariant derivatives $Y_i$ are defined by $Y_i=L_i+C_i$. The action (\ref{act1}) with the constraint (\ref{act0}) was studied extensively in \cite{ref}.

Following \cite{15,valv} we will derive in this article the supersymmetric analogue of the action (\ref{act1}). Let us summarize here the main results. Instead of the $SU(2)$ vector $\vec{C}=(C_i)$ we will have an $OSP(2,2)$ supervector  $A=(A_{\pm},W,C_i,B_{\pm})$. The  $5$ superfields  $C_{i}$, $B_{\pm}$ transform as a superspin $1$ multiplet under $OSP(2,1)$ while the remaining $3$ superfields $A_{\pm}$ and $W$ will transform as a superspin $1/2$ under $OSP(2,1)$.  All these superfields are elements of the algebra $Mat(2L+1,2L)$. We define the supersymmetric curvature by $F={\delta}A+A*A=(F_{\pm},f,c_i,b_{\pm})$ where the exterior derivative ${\delta}$ and the associative product $*$ are defined appropriately on forms. The action by analogy with (\ref{act1}) reads

\begin{eqnarray}
S_L[A]
&=&{\alpha}Str\triangleleft F*F+\beta STr(A*F-\frac{1}{3}A*A*A).\label{act2}
\end{eqnarray}
The first term is similar to the usual Yang-Mills action whereas the second term is a ( real-valued ) Chern-Simons-like contribution. $\alpha$ and $\beta$ are two real parameters. The Hodge triangle $\triangleleft $ is defined as the identity map between  one-forms and two-forms and thus $\triangleleft F$ should be considered as a one-form. This action will have the correct continuum limit provided we impose the following supersymmetric- and gauge-covariant conditions on the supergauge field $A$. The first condition is the supersymmetric anlogue of (\ref{act0}) defined by 

\begin{eqnarray}
[D_+,A_-]-[D_-,A_+]+\frac{1}{4}\{D_0,W\}+[A_+,A_-]+\frac{1}{4}W^2=0.\label{act3}
\end{eqnarray}
We will also impose the following supersymmetric constraints
\begin{eqnarray}
b_+=b_-=c_+=c_-=c_3=0.\label{act4}
\end{eqnarray}
These constraints will reduce the number of independent components of $A$ and $F$ from $8$ to $2$. $D_{\pm,0}$ are the generators of $OSP(2,2)$ in the complement of $OSP(2,1)$. The generators of $OSP(2,1)$ are denoted by $R_i$, $V_{\pm}$. The expressions of the curvatures $F_{\pm}$, $f$  and $c_i,b_{\pm}$ in terms of the gauge fields $A_{\pm},W$ and $C_i,B_{\pm}$ are given by 
\begin{eqnarray}
&&F_{\pm}=2[X_{\mp},Y_{\pm}]{\pm}2[X_{\pm},Y_3]+[X_0,Z_{\pm}]+2X_{\pm}~,~f=4\{Z_+,X_-\}-4\{Z_-,X_+\}+2X_0\nonumber\\
&&c_{\pm}={\mp}2\{X_{\pm},X_{\pm}\}-Y_{\pm}~,~c_3=2\{X_+,X_-\}-Y_3~,~b_{\pm}=[X_0,X_{\pm}]-Z_{\pm}.
\end{eqnarray}
The supercovariant derivatives $X_{\pm},X_0,Y_i$ and $Z_{\pm}$ are defined by $X_{\pm}=D_{\pm}+A_{\pm}$ , $X_0=D_0+W$,$Y_i=R_i+C_i$ and $~Z_{\pm}=V_{\pm}+B_{\pm}$.

In the rest of much of these notes we will go into the detail of the above construction following \cite{15,valv}. Motivated by this construction we introduce in section $4$ a novel fuzzy supersymmetric scalar action on the fuzzy supersphere. 

For other constructions of fuzzy supersymmetric gauge models   see \cite{iso}. The work \cite{17} is a numerical study of the type of supersymmetry which is involved in  IKKT models \cite{IKKT,IKKT1} so it is not the same as fuzzy SUSY. This paper is organized as follows 

\tableofcontents
\section{The continuum supersphere}
In this section we will follow \cite{Bal,14}.
\subsection{The Lie algebras $osp(2,1)$ and $osp(2,2)$}
We start with the $osp(2,1)$ Lie algebra. It consists of three even generators $R_{\pm}$ and $R_{3}$ and two odd generators $V_{\pm}$ with commutation and anticommutation relations
\begin{eqnarray} 
[R_{+},R_{-}]=2R_{3}~,~[R_3,R_{\pm}]=\pm R_{\pm}
\end{eqnarray}
and
\begin{eqnarray}
[R_3,V_{\pm}]=\pm\frac{1}{2}V_{\pm}~,~[R_{\pm},V_{\pm}]=0~,~[R_{\pm},V_{\mp}]=V_{\pm}.\label{2}
\end{eqnarray}
\begin{eqnarray}
\{V_{\pm},V_{\pm}\}=\pm \frac{1}{2}R_{\pm}~,~\{V_{\pm},V_{\mp}\}=- \frac{1}{2}R_{3}.\label{3}
\end{eqnarray}
The following notation is also useful $R_{\pm}={\Lambda}_1\pm i{\Lambda}_2$, $R_{3}={\Lambda}_3$, $V_{+}={\Lambda}_4$ and $V_{-}={\Lambda}_5$. The above commutation and anticommutation relations take now the following forms respectively
\begin{eqnarray}
[{\Lambda}_i,{\Lambda}_j]=i{\epsilon}_{ijk}{\Lambda}_k~,~[{\Lambda}_i,{\Lambda}_{\alpha}]=\frac{1}{2}({\sigma}_i)_{\beta \alpha}{\Lambda}_{\beta}~,~\{{\Lambda}_{\alpha},{\Lambda}_{\beta}\}=\frac{1}{2}(C{\sigma}_i)_{\alpha \beta }{\Lambda}_{i}.\label{834}
\end{eqnarray}
$i,j,k=1,2,3$ and $\alpha, \beta=4,5$. The charge conjugation $C$ is such that $C_{45}=-C_{54}=1$ and $C_{44}=C_{55}=0$. The most important point is that $V_{\pm}$ transform as an $SU(2)$ spinor. 

Let us also introduce $osp(2,2)$. We add two more odd generators $D_{\pm}$ and one even generator $D_0$ with the commutation and anticommutation relations
\begin{eqnarray}
[R_3,D_{\pm}]=\pm\frac{1}{2}D_{\pm}~,~[R_{\pm},D_{\pm}]=0~,~[R_{\pm},D_{\mp}]=D_{\pm}.\label{5}
\end{eqnarray}
\begin{eqnarray}
\{D_{\pm},D_{\pm}\}=\mp \frac{1}{2}R_{\pm}~,~\{D_{\pm},D_{\mp}\}= \frac{1}{2}R_{3}.\label{6}
\end{eqnarray}
\begin{eqnarray}
\{D_{\pm},V_{\pm}\}=0~,~\{D_{\pm},V_{\mp}\}= \pm \frac{1}{4}D_0.\label{7}
\end{eqnarray}
and
\begin{eqnarray}
[D_0 , R_i]=0~,~[D_0, V_{\pm}]=D_{\pm}~,~[D_0,D_{\pm}]=V_{\pm}.\label{8}
\end{eqnarray}
 Again we denote $D_{+}={\Lambda}_6$, $D_{-}={\Lambda}_7$ and $D_0={\Lambda}_8$. Then the  commutation and anticommutation relations  (\ref{5}) and (\ref{6}) take  the forms respectively
\begin{eqnarray}
[{\Lambda}_i,{\Lambda}_{\alpha}]=\frac{1}{2}({\sigma}_i)_{\beta \alpha}{\Lambda}_{\beta}~,~\{{\Lambda}_{\alpha},{\Lambda}_{\beta}\}=-\frac{1}{2}(C{\sigma}_i)_{\alpha \beta }{\Lambda}_{i}.\label{87g}
\end{eqnarray}
Here $\alpha, \beta=6,7$. Note that $D_{\pm}$ transform also as an $SU(2)$ spinor.  Equation (\ref{2}) and (\ref{5}) can be put in the form 
\begin{eqnarray}
[{\Lambda}_i,{\Lambda}_{\alpha}]=\frac{1}{2}(\tilde{\sigma}_i)_{\beta \alpha}{\Lambda}_{\beta}~
\end{eqnarray}
Equation (\ref{3}), (\ref{6}) and (\ref{7}) can be put together in the form
\begin{eqnarray}
\{{\Lambda}_{\alpha},{\Lambda}_{\beta}\}=\frac{1}{2}(\tilde{C}\tilde{\sigma}_i)_{\alpha \beta }{\Lambda}_{i}+\frac{1}{4}(\tilde{\epsilon}\tilde{C})_{\alpha \beta}{\Lambda}_8.
\end{eqnarray}
Equation (\ref{8}) takes the form
\begin{eqnarray}
[{\Lambda}_8,{\Lambda}_i]=0~,~[{\Lambda}_8,{\Lambda}_{\alpha}]=\tilde{\epsilon}_{\alpha \beta}{\Lambda}_{\beta}.
\end{eqnarray}
Here ( in the last three equations ) $\alpha,\beta=4,5,6,7$ and 
\begin{eqnarray}
\tilde{\sigma}_i=\bigg(\begin{array}{cc}
                   {\sigma}_i & 0 \\
            0 & {\sigma}_i  
                 \end{array}\bigg)~,~\tilde{C}=\bigg(\begin{array}{cc}
                   {C} & 0 \\
            0 & -{C}  
                 \end{array}\bigg)~,~\tilde{\epsilon}=\bigg(\begin{array}{cc}
                   0 & {\bf 1}_2 \\
            {\bf 1}_2 & 0  
                 \end{array}\bigg).
\end{eqnarray}
Also
\begin{eqnarray}
{\sigma}_1=\bigg(\begin{array}{cc}
                   0 & 1 \\
            1 & 0  
                 \end{array}\bigg)~,~{\sigma}_2=\bigg(\begin{array}{cc}
                   0 & -i \\
            i & 0  
                 \end{array}\bigg)~,~{\sigma}_3=\bigg(\begin{array}{cc}
                   1 & 0 \\
            0 & -1  
                 \end{array}\bigg).
\end{eqnarray}
\subsection{The supersphere}
The supersphere ${\bf S}^{(3,2)}$ ( with ordinary ${\bf S}^3$ as its even part ) is given by the points $\psi {\in}~{\bf C}^{(2,1)}$ which satisfy
\begin{eqnarray}
|\psi |^2=1.
\end{eqnarray}
We have $\psi =(z,\theta)=(z_1,z_2,\theta )$ and $\bar{\psi}=(\bar{z},\bar{\theta})=(z_1^+,z_2^+,\bar{\theta})$ and the norm is given by
\begin{eqnarray}
|\psi |^2\equiv\bar{\psi}{\psi}=\bar{z}z+\bar{\theta}{\theta}=|z_1|^2+|z_2|^2+\bar \theta \theta.
\end{eqnarray}
In above $z_1$, $z_2$ are complex variables and $\theta$, $\bar\theta $ are Grassmann numbers. The group manifold of $osp(2,1)$ is ${\bf S}^{(3,2)}$ in the same way that the group manifold of $su(2)$ is ${\bf S}^3$. Furthermore the supersphere ${\bf S}^{(2,2)}$ is an adjoint orbit of $OSP(2,1)$ in the same way that the sphere ${\bf S}^{2}$ is an adjoint orbit of $SU(2)$. In other words we must consider the supersymmetric Hopf fibration ${\bf S}^1{\longrightarrow}{\bf S}^{(3,2)}{\longrightarrow}{\bf S}^{(2,2)}$ by analogy with the Hopf fibration ${\bf S}^1{\longrightarrow}{\bf S}^{3}{\longrightarrow}{\bf S}^{2}$. We define thus the coordinates functions on ${\bf S}^{(2,2)}$ by the following functions on ${\bf S}^{3,2}$
\begin{eqnarray}
{\omega}_a(\psi)=\bar{\psi}{\Lambda}_a^{(\frac{1}{2})}{\psi}~,~a=1,...,5.
\end{eqnarray}
A point on ${\bf S}^{(2,2)}$ is given by the supervector $\omega =({\omega}_1,...,{\omega}_5)$.
${\Lambda}_a^{(\frac{1}{2})}$ are the generators of $OSP(2,1)$ in the $3-$dimensional fundamental representation characterized by the superspin $j=\frac{1}{2}$. It consists of the $SU(2)$ irreducible representations $\frac{1}{2}$ and $0$. The generators are given explicitly by
\begin{eqnarray}
{\Lambda}_i^{(\frac{1}{2})}=\frac{1}{2}\Bigg(\begin{array}{cc}
                   {\sigma}_i & 0 \\
            0 & 0  
                 \end{array}\Bigg)~,~{\Lambda}_4^{(\frac{1}{2})}=\frac{1}{2}\Bigg(\begin{array}{ccc}
                   0 & 0&-1 \\
            0 & 0&0\\
           0&-1&0  
                 \end{array}\Bigg)~,~{\Lambda}_5^{(\frac{1}{2})}=\frac{1}{2}\Bigg(\begin{array}{ccc}
                   0 & 0&0 \\
            0 & 0&-1\\
           1&0&0  
                 \end{array}\Bigg).\label{17}
\end{eqnarray}
Remark that under $\psi {\longrightarrow} \psi h$, $h=\exp(i\gamma)$ with $\gamma $ real numbers we have ${\omega}_a{\longrightarrow}{\omega}_a$. Thus the points $ \psi h$  on ${\bf S}^{(3,2)}$ correspond to the same point $\omega$ on ${\bf S}^{(2,2)}$. This shows that $OSP(2,1)$ is a principal $U(1)$ bundle over the coset space $OSP(2,1)/U(1)$ which is diffeomorphic to the  sphere ${\bf S}^{(2,2)}$. We can compute the explicit expressions
\begin{eqnarray}
{\omega}_i=\frac{1}{2}\bar{z}{\sigma}_iz~,~{\omega}_4=-\frac{1}{2}({z}_1^+\theta +z_2\bar{\theta})~,~{\omega}_5=\frac{1}{2}(-{z}_2^+\theta+z_1\bar{\theta}).
\end{eqnarray}
By using these equations we can immediately compute
\begin{eqnarray}
{\omega}_i^2+C_{\alpha \beta}{\omega}_{\alpha}{\omega}_{\beta}=\frac{1}{4}.\label{s22}
\end{eqnarray}
This is the defining equation of the supersphere ${\bf S}^{(2,2)}$. We define the grade adjoint $++$ by $z^{++}_i={z}_i^{+}$, ${\theta}^{++}=\bar{\theta}$ and $\bar{\theta}^{++}=-{\theta}$ and by the requirement that $(AB)^{++}=(-1)^{d_Ad_B}B^{++}A^{++}$ where $d_A$ and $d_B$ are the degrees of $A$ and $B$ respectively. For an even object the degree is equal $0$ while for an odd object the degree is equal to $1$. Hence we have the reality conditions 
\begin{eqnarray}
{\omega}_i^{++}\equiv{\omega}_i^+={\omega}_i~,~{\omega}^{++}_{\alpha}=-C_{\alpha \beta}{\omega}_{\beta}~,~\alpha,\beta=4,5. \label{grade+}
\end{eqnarray}
The action of the group $OSP(2,2)$ on ${\bf S}^{(2,2)}$ preserves (\ref{s22}) and (\ref{grade+}) but it is not the same  as the adjoint action of the group $OSP(2,1)$. This is because the Lie algebra $osp(2,1)$ is not invariant under the action of the generators ${\Lambda}_6$, ${\Lambda}_7$ and ${\Lambda}_8$ of $OSP(2,2)$.
 Let us define the $OSP(2,2)$ coordinates functions

\begin{eqnarray}
{\Omega}_a(\psi)=\bar{\psi}{\Lambda}_a^{(\frac{1}{2})}{\psi}~,~a=1,...,8.
\end{eqnarray}
They define an $OSP(2,2)$ vector. We will have the following extra generators ( in addition to (\ref{17}) ) in the $3-$dimensional fundamental representation $j=\frac{1}{2}$ of $OSP(2,2)$ 

\begin{eqnarray}
{\Lambda}_8^{(\frac{1}{2})}=\Bigg(\begin{array}{cc}
                   {\bf 1}_2 & 0 \\
            0 & 2  
                 \end{array}\Bigg)~,~{\Lambda}_6^{(\frac{1}{2})}=\frac{1}{2}\Bigg(\begin{array}{ccc}
                   0 & 0&1 \\
            0 & 0&0\\
           0&-1&0  
                 \end{array}\Bigg)~,~{\Lambda}_7^{(\frac{1}{2})}=\frac{1}{2}\Bigg(\begin{array}{ccc}
                   0 & 0&0 \\
            0 & 0&1\\
           1&0&0  
                 \end{array}\Bigg).\label{uuz}
\end{eqnarray}
Explicitly we have
\begin{eqnarray}
{\Omega}_8=2-\bar{z}z~,~{\Omega}_6=\frac{1}{2}({z}_1^+\theta -z_2\bar{\theta})~,~{\Omega}_7=\frac{1}{2}({z}_2^+\theta+z_1\bar{\theta})
\end{eqnarray}
\begin{eqnarray}
{\Omega}_8^{++}\equiv{\Omega}_8^+={\Omega}_8~,~{\Omega}^{++}_{\alpha}=C_{\alpha \beta }{\Omega}_{\beta}~,\alpha,\beta =6,7. \label{grade+2}
\end{eqnarray}
We can immediately compute
\begin{eqnarray}
-\frac{1}{4}{\Omega}_8^2+\tilde{C}_{\alpha \beta}{\Omega}_{\alpha}{\Omega}_{\beta}=-\frac{1}{4}~,\alpha,\beta =6,7.\label{s22-2}
\end{eqnarray}
By adding (\ref{s22}) ( with the substitutions ${\omega}_i{\longrightarrow}{\Omega}_i$ and ${\omega}_{\alpha}{\longrightarrow}{\Omega}_{\alpha}~,~\alpha=4,5$ ) and (\ref{s22-2}) we get the $OSP(2,2)$ Casimir
\begin{eqnarray}
{\Omega}_i^2-\frac{1}{4}{\Omega}_8^2+\tilde{C}_{\alpha \beta}{\Omega}_{\alpha}{\Omega}_{\beta}=0.\label{s22-3}
\end{eqnarray}
Let us also compute the following
\begin{eqnarray}
{\omega}_3{\omega}_4=(\frac{1}{2}|z_1|^2-\frac{1}{2}|z_2|^2){\omega}_4=-\frac{1}{4}(|z_1|^2z_1^+\theta +|z_1|^2z_2\bar{\theta}-|z_2|^2z_1^+\theta-|z_2|^2z_2\bar{\theta})
\end{eqnarray}
\begin{eqnarray}
({\omega}_1+i{\omega}_2){\omega}_5=(z_1^+z_2){\omega}_5=\frac{1}{2}(|z_1|^2z_2\bar{\theta}-|z_2|^2z_1^+\theta).
\end{eqnarray}
Hence by using also $\sqrt{{\omega}_i^2}=\frac{1}{2}\bar{z}z$ we obtain
\begin{eqnarray}
{\Omega}_6=-\frac{1}{\sqrt{{\omega}_i^2}}({\omega}_3{\omega}_4+({\omega}_1+i{\omega}_2){\omega}_5).\label{om6}
\end{eqnarray}
By using ${\Omega}_6^{++}={\Omega}_7$, ${\omega}_4^{++}=-{\omega}_5$, ${\omega}_5^{++}={\omega}_4$, ${\omega}_i^{++}={\omega}_i$ and $(i)^{++}=-i$ we obatin 
\begin{eqnarray}
{\Omega}_7=\frac{1}{\sqrt{{\omega}_i^2}}({\omega}_3{\omega}_5-({\omega}_1-i{\omega}_2){\omega}_4).\label{om7}
\end{eqnarray}
Finally by using again $\sqrt{{\omega}_i^2}=\frac{1}{2}\bar{z}z$ we obtain
\begin{eqnarray}
{\Omega}_8=2-2\sqrt{{\omega}_i^2}.
\end{eqnarray}
\subsection{Laplacians}
Define $n_i=2R{\omega}_i,n_{\alpha}=2R{\omega}_{\alpha}$. Then
\begin{eqnarray}
n_i^2+C_{\alpha \beta}n_{\alpha}n_{\beta}=R^2.
\end{eqnarray}
The delta function ${\delta}(n_i^2+C_{\alpha \beta}n_{\alpha}n_{\beta}-R^2)$ will have an expansion of the general form ${\delta}(n_i^2+C_{\alpha \beta}n_{\alpha}n_{\beta}-R^2)={\delta}(n_i^2-R^2)+C_{\alpha \beta}n_{\alpha}n_{\beta}X$ where $X$ is given by
\begin{eqnarray}
X=\frac{1}{2}\big[\frac{d}{dn_5dn_4}{\delta}(n_i^2+C_{\alpha \beta}n_{\alpha}n_{\beta}-R^2)\big]_{n_4=n_5=0}=\frac{d{\delta}(n_i^2-R^2)}{dn_i^2}.
\end{eqnarray}
Thus 
\begin{eqnarray}
 {\delta}(n_i^2+C_{\alpha \beta}n_{\alpha}n_{\beta}-R^2)=\frac{1}{2R}{\delta}(r-R)+\frac{n_4n_5}{2rR}\frac{d{\delta}(r-R)}{dr}~,~n_i^2=r^2.
\end{eqnarray}
In above we have assumed that $r{\geq}0$. A scalar superfield $\Phi$ on ${\bf S}^{(2,2)}$ has an expansion of the form ( with $\alpha ,\beta =4,5$ )
\begin{eqnarray}
\Phi={\phi}_0+C_{\alpha \beta}{\psi}_{\alpha}{n}_{\beta}+{\phi}_1C_{\alpha \beta}n_{\alpha }n_{\beta}.
\end{eqnarray}
${\phi}_0\equiv{\phi}_0(n_i)$ and ${\phi}_1\equiv{\phi}_1(n_i)$ are scalar functions while ${\psi}$ is a  Majorana spinor field with two components Grassman functions ${\psi}_{\alpha}\equiv{\psi}_{\alpha}(n_i)$. In the terminology of supesymmetry in $4-$dimensional Minkowski spacetime the field ${\phi}_0$ is the $D-$term of the superfield ${\Phi}$ while the field  ${\phi}_1$ is the $F-$term of the superfield. The integral of this superfield over the supersphere is defined by  ( with $d{\Omega}$ denoting the solid angle )
\begin{eqnarray}
I(\Phi)=\int r^2dr d{\Omega}dn_4dn_5{\delta}(n_i^2+C_{\alpha \beta}n_{\alpha}n_{\beta}-R^2)\Phi.
\end{eqnarray}
Since the volum form and the delta function are invariant under the $OSP(2,1)$ action the integral should be invariant under the susy action on $\Phi$.
A straightforward calculation ( using also $\int dn_4=\int dn_5=0$ and $\int dn_4dn_5n_4n_5=-1$ ) we obtain
\begin{eqnarray}
I(\Phi)=\int d{\Omega}\bigg[\frac{d}{dr}(\frac{r}{2R}{\phi}_0)-R{\phi}_1\bigg]_{r=R}.
\end{eqnarray}
Thus as in the case of supersymmetry in $4$ dimensions the integral depends only on the $D-$ and $F-$terms of the superfield.
The $OSP(2,1)$ and $OSP(2,2)$ Laplacians ( by inspection of equations (\ref{s22}), (\ref{s22-2}) and (\ref{s22-3}) ) are given respectively by the equations
\begin{eqnarray}
K_{2,1}&=&{\Lambda}_i^2+C_{\alpha \beta}{\Lambda}_{\alpha}{\Lambda}_{\beta}\nonumber\\
K_{2,2}&=&{\Lambda}_i^2-\frac{1}{4}{\Lambda}_8^2+\tilde{C}_{\alpha \beta}{\Lambda}_{\alpha}{\Lambda}_{\beta}.
\end{eqnarray}
The Laplacian on ${\bf S}^{(2,2)}$ is given by
\begin{eqnarray}
{\Delta}=K_{2,1}-K_{2,2}=\frac{1}{4}{\Lambda}_8^2+{\Lambda}_6{\Lambda}_7-{\Lambda}_7{\Lambda}_6.
\end{eqnarray}
The action for the scalar superfield $\Phi$ is given by
\begin{eqnarray}
S=I({\Phi}^{++}{\Delta}{\Phi})=\int r^2dr d{\Omega}dn_4dn_5{\delta}(n_i^2+C_{\alpha \beta}n_{\alpha}n_{\beta}-R^2){\Phi}^{++}{\Delta}{\Phi}.
\end{eqnarray}
\subsection{Scalar action}
In the calculation of the above action we need the $D-$ and $F-$terms of the superfield ${\Phi}^{++}{\Delta}{\Phi}$. Because of the constraint the superfield $\Phi$ can be rewritten in the form $\Phi={\phi}_2+C_{\alpha \beta}{\psi}_{\alpha}n_{\beta}$ where ${\phi}_2={\phi}_0+{\phi}_1(R^2-r^2)$. Hence the $D-$term of the superfield ${\Phi}^{++}{\Delta}{\Phi}$ is
\begin{eqnarray}
[{\Phi}^{++}{\Delta}{\Phi}]_0={\phi}_2^{++}{\Delta}{\phi}_2={\phi}_2{\Delta}{\phi}_2.
\end{eqnarray}
The F$-$term will be extracted from
\begin{eqnarray}
 [{\Phi}^{++}{\Delta}{\Phi}]_1&=&
(C_{\alpha \beta}{\psi}_{\alpha}n_{\beta})^{++}{\Delta}(C_{\alpha \beta}{\psi}_{\alpha}n_{\beta})
=(C_{\alpha \beta}{\psi}_{\alpha}n_{\beta}){\Delta}(C_{\alpha \beta}{\psi}_{\alpha}n_{\beta}).
\end{eqnarray}
In above we have assumed that the superfield $\Phi$ is real and hence ${\Phi}^{++}=\Phi$ or equivalently ${\phi}_0^{++}={\phi}_0^+={\phi}_0$,  ${\phi}_1^{++}={\phi}_1^+={\phi}_1$ and ${\psi}_{\alpha}^{++}=-C_{\alpha \beta}{\psi}_{\beta}$.  We have also assumed that cross terms are linear in $n_{\alpha}$ which we will show.

\paragraph{The D$-$term :} First we calculate the $D-$component. The action of the generators ${\Lambda}_6=D_+$ and ${\Lambda}_7=D_{-}$ on ${\phi}_0$ is defined by \footnote{Take the case of the ordinary generators of $SU(2)$ denoted here by ${\Lambda}_i=R_i$. We know that ${\Lambda}_i{\phi}_0\equiv ({\cal R}_i{\phi}_0)(\vec{n})=-i{\epsilon}_{ijk}n_j{\partial}_k{\phi}_0$. This can be put in the form ${\Lambda}_i{\phi}_0=({\cal R}_in_j){\partial}_j{\phi}_0$.}
\begin{eqnarray}
{\Lambda}_{6}{\phi}_0\equiv ({\cal D}_{+}n_i){\partial}_i{\phi}_0&=&-\frac{1}{2}\bigg[({\sigma}_i)_{ 66 }n_6+({\sigma}_i)_{7 6}n_7\bigg]{\partial}_i{\phi}_0=-\frac{1}{2}\big[n_6{\partial}_3+n_7{\partial}_+\big]{\phi}_0\nonumber\\
{\Lambda}_{7}{\phi}_0\equiv ({\cal D}_{-}n_i){\partial}_i{\phi}_0&=&-\frac{1}{2}\bigg[({\sigma}_i)_{ 67 }n_6+({\sigma}_i)_{7 7}n_7\bigg]{\partial}_i{\phi}_0=-\frac{1}{2}\big[n_6{\partial}_{-}-n_7{\partial}_3\big]{\phi}_0.\label{sime}
\end{eqnarray}
These equations are consistent with the commutation relations $[{\Lambda}_{\alpha},{\Lambda}_i]=-\frac{1}{2}({\sigma}_i)_{\beta \alpha}{\Lambda}_{\beta}$ where $\alpha, \beta=6,7$. Let us also say that the operators ${\cal D}_{\pm}$ correspond to the generators $D_{\pm}$ in the adjoint representation of $OSP(2,2)$. Furthermore ${\partial}_{\pm}={\partial}_1\pm i{\partial}_2$ and  $n_6$, $n_7$ are given by $n_6=2R{\Omega}_6$, $n_7=2R{\Omega}_7$ and hence we must have from (\ref{om6}) and (\ref{om7}) the expressions

\begin{eqnarray}
{n}_6=-\frac{1}{r}({n}_3{n}_4+({n}_1+i{n}_2){n}_5)~,~
{n}_7=\frac{1}{r}({n}_3{n}_5-({n}_1-i{n}_2){n}_4).\label{765}
\end{eqnarray} 
Remark that ${\Lambda}_{6}{\phi}_0$ and ${\Lambda}_{7}{\phi}_0$ are odd and hence a second action of ${\Lambda}_6$ and ${\Lambda}_7$ will involve anticommutation relations instead of commutation relations. We have
\begin{eqnarray}
{\Lambda}_7{\Lambda}_{6}{\phi}_0&=&-\frac{1}{2}\bigg[{\Lambda}_7(n_6{\partial}_3{\phi}_0)+{\Lambda}_7(n_7{\partial}_+{\phi}_0)\bigg]\nonumber\\
&=&-\frac{1}{2}\bigg[({\Lambda}_7n_6){\partial}_3{\phi}_0-n_6({\Lambda}_7{\partial}_3{\phi}_0)+({\Lambda}_7n_7){\partial}_+{\phi}_0-n_7({\Lambda}_7{\partial}_+{\phi}_0)\bigg]\nonumber\\
{\Lambda}_6{\Lambda}_{7}{\phi}_0&=&-\frac{1}{2}\bigg[{\Lambda}_6(n_6{\partial}_{-}{\phi}_0)-{\Lambda}_6(n_7{\partial}_3{\phi}_0)\bigg]\nonumber\\
&=&-\frac{1}{2}\bigg[({\Lambda}_6n_6){\partial}_{-}{\phi}_0-n_6({\Lambda}_6{\partial}_{-}{\phi}_0)-({\Lambda}_6n_7){\partial}_3{\phi}_0+n_7({\Lambda}_6{\partial}_3{\phi}_0)\bigg]\nonumber\\
\end{eqnarray}
The quantities $({\Lambda}_{\alpha}{\partial}_3{\phi}_0)$  and $({\Lambda}_{\alpha}{\partial}_{\pm}{\phi}_0)$ will be given by similar expressions to (\ref{sime}). From the anticommutation relations $\{D_{\pm},D_{\pm}\}={\mp}\frac{1}{2}R_{\pm}$ and $\{D_{\pm},D_{\mp}\}=\frac{1}{2}R_{3}$ we have 
\begin{eqnarray}
&&{\Lambda}_6n_6={\cal D}_+n_6=-\frac{1}{2}n_+~,~{\Lambda}_6n_7={\cal D}_+n_7=\frac{1}{2}n_3~,~n_+=n_1+in_2\nonumber\\
&&{\Lambda}_7n_6={\cal D}_-n_6=\frac{1}{2}n_3~,~{\Lambda}_7n_7={\cal D}_-n_7=+\frac{1}{2}n_-~,~n_-=n_1-in_2.
\end{eqnarray}
Note here that the odd coordinates associated with ${\Lambda}_{6,7}$ will always be denoted by $n_{6,7}$ although we will denote sometimes the operators  ${\Lambda}_{6}$ and ${\Lambda}_{7}$   by $D_+$ and $D_-$ respectively. So $n_+$ and $n_-$ are always bosonic coordinates associated with ${\Lambda}_+={\Lambda}_1+i{\Lambda}_2$ and ${\Lambda}_-={\Lambda}_1-i{\Lambda}_2$. We compute ( with ${\cal L}_3=i(n_1{\partial}_2-n_2{\partial}_1)$ ) 
\begin{eqnarray}
&&{\Lambda}_7{\Lambda}_6{\phi}_0=-\frac{1}{4}\big[(+n_i{\partial}_i+{\cal L}_3){\phi}_0-n_6n_7{\partial}^2{\phi}_0\big]\nonumber\\
&&{\Lambda}_6{\Lambda}_7{\phi}_0=-\frac{1}{4}\big[(-n_i{\partial}_i+{\cal L}_3){\phi}_0+n_6n_7{\partial}^2{\phi}_0\big].
\end{eqnarray}
Thus
\begin{eqnarray}
\big({\Lambda}_6{\Lambda}_7-{\Lambda}_7{\Lambda}_6\big)({\phi}_0)=\frac{1}{2}\big[n_i{\partial}_i{\phi}_0-n_6n_7{\partial}^2{\phi}_0\big].
\end{eqnarray}
Similarly ${\Lambda}_8({\phi}_0)={\cal D}_0(n_i){\partial}_i{\phi}_0=0$ since $[{\Lambda}_8,R_i]=[D_0,R_i]=0$. Hence
\begin{eqnarray}
{\phi}_0{\Delta}({\phi}_0)=\frac{1}{2}{\phi}_0\big[n_i{\partial}_i{\phi}_0-n_6n_7{\partial}^2{\phi}_0\big]&=&\frac{1}{2}{\phi}_0\bigg[r{\partial}_r{\phi}_0+\frac{R^2-r^2}{2}{\partial}^2{\phi}_0\bigg]\nonumber\\
&=&\frac{1}{2}{\phi}_0\bigg[\frac{R^2-r^2}{2}{\partial}_r^2+\frac{R^2}{r}{\partial}_r+\frac{1}{2}(\frac{R^2}{r^2}-1){\cal L}_a^2\bigg]{\phi}_0.
\end{eqnarray}
In above we have used the results $n_6n_7=-n_4n_5=-\frac{R^2-r^2}{2}$ and ${\partial}^2={\partial}_r^2+\frac{2}{r}{\partial}_r+\frac{{\cal L}_a^2}{r^2}$. Finally we get
\begin{eqnarray}
\frac{d}{dr}\bigg[\frac{r}{2R}{\phi}_0{\Delta}{\phi}_0\bigg]_{r=R}=\frac{R}{4}\bigg[\bigg(\frac{d{\phi}_0}{dr}\bigg)^2\bigg]_{r=R}-\frac{1}{4R}{\phi}_0{\cal L}_a^2{\phi}_0.
\end{eqnarray}
The corresponding action is
\begin{eqnarray}
I_0=\int d{\Omega}\frac{d}{dr}\bigg[\frac{r}{2R}{\phi}_0{\Delta}{\phi}_0\bigg]_{r=R}=\frac{R}{4}\int d{\Omega}\bigg(\frac{d{\phi}_0}{dr}\bigg)^2+\frac{1}{4R}\int d{\Omega} ({\cal L}_a{\phi}_0)^2.
\end{eqnarray}
The full action coming from the D$-$term is obtained from above by replacing ${\phi}_0$ with ${\phi}_2$. We get
\begin{eqnarray}
I_D&=&\int d{\Omega}\frac{d}{dr}\bigg[\frac{r}{2R}{\phi}_2{\Delta}{\phi}_2\bigg]_{r=R}=\frac{R}{4}\int d{\Omega}\bigg(\frac{d{\phi}_2}{dr}\bigg)^2+\frac{1}{4R}\int d{\Omega} ({\cal L}_a{\phi}_2)^2\nonumber\\
&=&\int d{\Omega}\frac{d}{dr}\bigg[\frac{r}{2R}{\phi}_2{\Delta}{\phi}_2\bigg]_{r=R}=\frac{R}{4}\int d{\Omega}\bigg(\frac{d{\phi}_0}{dr}-2R{\phi}_1\bigg)^2+\frac{1}{4R}\int d{\Omega} ({\cal L}_a{\phi}_0)^2.
\end{eqnarray}

\paragraph{The $F-$term :}
Now we have ( with $D_-={\Lambda}_7$, $D_+={\Lambda}_6$ and $\alpha=4,5$)
\begin{eqnarray}
{\Lambda}_6({\psi}_{\alpha})={\cal D}_+(n_i){\partial}_i{\psi}_{\alpha}&=&\frac{1}{2}({\cal D}_+n_+){\partial}_-{\psi}_{\alpha}+\frac{1}{2}({\cal D}_+n_-){\partial}_+{\psi}_{\alpha}+{\cal D}_+(n_3){\partial}_3{\psi}_{\alpha}\nonumber\\
&=&-\frac{1}{2}n_7{\partial}_+{\psi}_{\alpha}-\frac{1}{2}n_6{\partial}_3{\psi}_{\alpha}\nonumber\\
{\Lambda}_7({\psi}_{\alpha})={\cal D}_-(n_i){\partial}_i{\psi}_{\alpha}&=&\frac{1}{2}({\cal D}_-n_+){\partial}_-{\psi}_{\alpha}+\frac{1}{2}({\cal D}_-n_-){\partial}_+{\psi}_{\alpha}+{\cal D}_-(n_3){\partial}_3{\psi}_{\alpha}\nonumber\\
&=&-\frac{1}{2}n_6{\partial}_-{\psi}_{\alpha}+\frac{1}{2}n_7{\partial}_3{\psi}_{\alpha}.
\end{eqnarray}
In above we have also used the fact ( which we can check from the commutation relations $[D_{\mp},R_{\pm}]=-D_{\pm}$, $[D_{\pm},R_{\pm}]=0$ and $[D_{\pm},R_3]={\mp}\frac{1}{2}D_{\pm}$ ) that
\begin{eqnarray}
{\cal D}_-(n_+)=-n_6~,{\cal D}_+(n_-)=-n_7~,{\cal D}_+(n_+)={\cal D}_-(n_-)=0~,{\cal D}_+(n_3)=-\frac{1}{2}n_6~,~{\cal D}_-(n_3)=\frac{1}{2}n_7
\end{eqnarray}
We will also need ( from the anticommutation relations $\{D_{\pm},V_{\pm}\}=0$ and $\{D_{\pm},V_{\mp}\}={\pm}\frac{1}{4}D_0$ with $D_0={\Lambda}_8$ ) the actions
\begin{eqnarray}
{\cal D}_+(n_5)=\frac{1}{4}n_8~,{\cal D}_+(n_4)=0~,{\cal D}_-(n_5)=0~,{\cal D}_-(n_4)=-\frac{1}{4}n_8.
\end{eqnarray}
The even coordinate $n_8$ is defined by $n_8=2R{\Omega}_8=2(2R-r)$. Next
\begin{eqnarray}
{\Lambda}_6(C_{\alpha \beta}{\psi}_{\alpha}n_{\beta})={\Lambda}_6({\psi}_4n_5-{\psi}_5n_4)&=&{\Lambda}_6({\psi}_4).n_5-{\psi}_4{\cal D}_+(n_5)-{\Lambda}_6({\psi}_5).n_4+{\psi}_5{\cal D}_+(n_4)\nonumber\\
&=&{\Lambda}_6({\psi}_4).n_5-\frac{1}{4}{\psi}_4n_8-{\Lambda}_6({\psi}_5).n_4\nonumber\\
&=&-\frac{1}{4}{\psi}_4n_8+\frac{1}{2}n_7n_5{\partial}_+{\psi}_4+\frac{1}{2}n_6n_5{\partial}_3{\psi}_4-\frac{1}{2}n_7n_4{\partial}_+{\psi}_5-\frac{1}{2}n_6n_4{\partial}_3{\psi}_5\nonumber\\
{\Lambda}_7(C_{\alpha \beta}{\psi}_{\alpha}n_{\beta})={\Lambda}_7({\psi}_4n_5-{\psi}_5n_4)&=&{\Lambda}_7({\psi}_4).n_5-{\psi}_4{\cal D}_-(n_5)-{\Lambda}_7({\psi}_5).n_4+{\psi}_5{\cal D}_-(n_4)\nonumber\\
&=&{\Lambda}_7({\psi}_4).n_5-{\Lambda}_7({\psi}_5).n_4-\frac{1}{4}{\psi}_5n_8\nonumber\\
&=&-\frac{1}{4}{\psi}_5n_8+\frac{1}{2}n_6n_5{\partial}_-{\psi}_4-\frac{1}{2}n_7n_5{\partial}_3{\psi}_4-\frac{1}{2}n_6n_4{\partial}_-{\psi}_5+\frac{1}{2}n_7n_4{\partial}_3{\psi}_5.\nonumber\\
\end{eqnarray}
By using now $n_7n_5=-\frac{1}{r}n_-n_4n_5$, $n_7n_4=n_6n_5=-\frac{1}{r}n_3n_4n_5$ and $n_6n_4=\frac{1}{r}n_+n_4n_5$ we obtain
\begin{eqnarray}
{\Lambda}_6(C_{\alpha \beta}{\psi}_{\alpha}n_{\beta})&=&-\frac{1}{4}{\psi}_4n_8-\frac{1}{2r}(n_-{\partial}_+{\psi}_4+n_3{\partial}_3{\psi}_4-n_3{\partial}_+{\psi}_5+n_+{\partial}_3{\psi}_5)n_4n_5\nonumber\\
{\Lambda}_7(C_{\alpha \beta}{\psi}_{\alpha}n_{\beta})
&=&-\frac{1}{4}{\psi}_5n_8-\frac{1}{2r}(n_3{\partial}_-{\psi}_4-n_-{\partial}_3{\psi}_4+n_+{\partial}_-{\psi}_5+n_3{\partial}_3{\psi}_5)n_4n_5.\label{789}
\end{eqnarray}
We need now to compute the following ( using also ${\cal D}_{\pm}(n_8)=-n_{4,5}$ )
\begin{eqnarray}
{\Lambda}_6({\psi}_{\alpha}n_8)&=&\frac{1}{2}{\cal D}_+(n_+){\partial}_-{\psi}_{\alpha}.n_8+\frac{1}{2}{\cal D}_+(n_-){\partial}_+{\psi}_{\alpha}.n_8+{\cal D}_+(n_3){\partial}_3{\psi}_{\alpha}.n_8-{\psi}_{\alpha}{\cal D}_+(n_8)\nonumber\\
&=&-\frac{1}{2}n_7{\partial}_+{\psi}_{\alpha}.n_8-\frac{1}{2}n_6{\partial}_3{\psi}_{\alpha}.n_8+{\psi}_{\alpha}n_4\nonumber\\
{\Lambda}_7({\psi}_{\alpha}n_8)&=&\frac{1}{2}{\cal D}_-(n_+){\partial}_-{\psi}_{\alpha}.n_8+\frac{1}{2}{\cal D}_-(n_-){\partial}_+{\psi}_{\alpha}.n_8+{\cal D}_-(n_3){\partial}_3{\psi}_{\alpha}.n_8-{\psi}_{\alpha}{\cal D}_-(n_8)\nonumber\\
&=&-\frac{1}{2}n_6{\partial}_-{\psi}_{\alpha}.n_8+\frac{1}{2}n_7{\partial}_3{\psi}_{\alpha}.n_8+{\psi}_{\alpha}n_5.
\end{eqnarray}

Next step is to compute the following action
\begin{eqnarray}
{\Lambda}_7\bigg(\frac{1}{2r}(n_-{\partial}_+{\psi}_4+n_3{\partial}_3{\psi}_4-n_3{\partial}_+{\psi}_5+n_+{\partial}_3{\psi}_5)n_4n_5\bigg)&=&-\frac{1}{8r}n_5n_8\big[n_-{\partial}_+{\psi}_4+n_3{\partial}_3{\psi}_4-n_3{\partial}_+{\psi}_5\nonumber\\
&+&n_+{\partial}_3{\psi}_5\big]\nonumber\\
{\Lambda}_6\bigg(\frac{1}{2r}(n_3{\partial}_-{\psi}_4-n_-{\partial}_3{\psi}_4+n_+{\partial}_-{\psi}_5+n_3{\partial}_3{\psi}_5)n_4n_5\bigg)&=&-\frac{1}{8r}n_4n_8(n_3{\partial}_-{\psi}_4-n_-{\partial}_3{\psi}_4+n_+{\partial}_-{\psi}_5\nonumber\\
&+&n_3{\partial}_3{\psi}_5).
\end{eqnarray}
This action corresponds to ${\Lambda}_{6,7}$ acting on the factor $n_4n_5$. The action of ${\Lambda}_{6,7}$ on the other terms leads to products which involve $n_4n_5$ and $n_{6,7}$ and hence they are zero by (\ref{765}). 

We now compute in a straightforward way
\begin{eqnarray}
&&n_5{\Lambda}_6{\Lambda}_7(C_{\alpha \beta}{\psi}_{\alpha}n_{\beta})=\frac{n_4n_5n_8}{8r}\bigg[(n_-{\partial}_+-n_+{\partial}_-){\psi}_5-(n_3{\partial}_--n_-{\partial}_3){\psi}_4-\frac{2r}{n_8}{\psi}_5\bigg]\nonumber\\
&&n_4{\Lambda}_6{\Lambda}_7(C_{\alpha \beta}{\psi}_{\alpha}n_{\beta})=\frac{n_4n_5n_8}{8r}(n_3{\partial}_+-n_+{\partial}_3){\psi}_5.
\end{eqnarray}

\begin{eqnarray}
&&n_4{\Lambda}_7{\Lambda}_6(C_{\alpha \beta}{\psi}_{\alpha}n_{\beta})=\frac{n_4n_5n_8}{8r}\bigg[(-n_+{\partial}_-+n_-{\partial}_+){\psi}_4-(n_3{\partial}_+-n_+{\partial}_3){\psi}_5+\frac{2r}{n_8}{\psi}_4\bigg]\nonumber\\
&&n_5{\Lambda}_7{\Lambda}_6(C_{\alpha \beta}{\psi}_{\alpha}n_{\beta})=\frac{n_4n_5n_8}{8r}(n_3{\partial}_--n_-{\partial}_3){\psi}_4.
\end{eqnarray}
So ( with ${\cal L}_3=\frac{1}{2}(n_+{\partial}_--n_-{\partial}_+)$, ${\cal L}_{\pm}={\cal L}_1\pm i{\cal L}_2=\mp (n_{\pm}{\partial}_3-n_3{\partial}_{\pm})$ and ${\psi}=({\psi}_1,{\psi}_2)$) 
\begin{eqnarray}
(C_{\alpha \beta}{\psi}_{\alpha}n_{\beta})({\Lambda}_6{\Lambda}_7-{\Lambda}_7{\Lambda}_6)(C_{\alpha \beta}{\psi}_{\alpha}n_{\beta})&=&\frac{n_4n_5n_8}{4r}\bigg[{\psi}_4{\cal L}_-{\psi}_4-{\psi}_4{\cal L}_3{\psi}_5-{\psi}_5{\cal L}_3{\psi}_4-{\psi}_5{\cal L}_+{\psi}_5\nonumber\\
&-&\frac{r}{n_8}{\psi}_4{\psi}_5+\frac{r}{n_8}{\psi}_5{\psi}_4\bigg]\nonumber\\
&=&-\frac{C_{\alpha \beta}n_{\alpha}n_{\beta}n_8}{8r}{\psi}^T({\sigma}_a{\cal L}_a+\frac{r}{n_8})(C\psi).
\end{eqnarray}
The full result is then
\begin{eqnarray}
(C_{\alpha \beta}{\psi}_{\alpha}n_{\beta}){\Delta}(C_{\alpha \beta}{\psi}_{\alpha}n_{\beta})
&=&-\frac{C_{\alpha \beta}n_{\alpha}n_{\beta}n_8}{8r}{\psi}^T({\sigma}_a{\cal L}_a+2\frac{r}{n_8})(C\psi).
\end{eqnarray}
The contribution of this F$-$term to the action is given by
\begin{eqnarray}
I_F=-R\int d{\Omega}\bigg[-\frac{n_8}{8r}{\psi}^T({\sigma}_a{\cal L}_a+2\frac{r}{n_8})(C\psi)]|_{r=R}=\frac{R}{4}\int d{\Omega} {\psi}^T({\sigma}_a{\cal L}_a+1)(C\psi).
\end{eqnarray}
The total action 
\begin{eqnarray}
I
&=&\frac{R}{4}\int d{\Omega}\bigg(\frac{d{\phi}_0}{dr}-2R{\phi}_1\bigg)^2+\frac{1}{4R}\int d{\Omega} ({\cal L}_a{\phi}_0)^2+\frac{R}{4}\int d{\Omega} {\psi}^T({\sigma}_a{\cal L}_a+1)(C\psi).
\end{eqnarray}
\section{The fuzzy supersphere}
We consider the irreducible representation with $OSP(2,1)$ superspin equal $L$. This representation consists of the direct sum of the  $SU(2)$ representations with spins $L$ and $L-\frac{1}{2}$.  Let ${\cal L}(L,L)$ be the space of linear operators from the corresponding representation space into itself. The action of the superalgebra $OSP(2,2)$ on ${\cal L}(L,L)$ is described by the operators 
\begin{eqnarray}
&&{R}_i=\bigg(\begin{array}{cc}
                   {R}_i^{(L)} & 0 \\
            0 & {R}_i^{(L-\frac{1}{2})}  
                 \end{array}\bigg)~,~{V}_{\alpha}=\bigg(\begin{array}{cc}
                   0 & V_{\alpha}^{(L,L-\frac{1}{2})} \\
            V_{\alpha}^{(L-\frac{1}{2},L)} & 0
                 \end{array}\bigg)\nonumber\\
&&D_0=\bigg(\begin{array}{cc}
                   2L & 0 \\
            0 & 2L+1  
                 \end{array}\bigg)~,~{D}_{\alpha}=\bigg(\begin{array}{cc}
                   0 & -V_{\alpha}^{(L,L-\frac{1}{2})} \\
            V_{\alpha}^{(L-\frac{1}{2},L)} & 0
                 \end{array}\bigg).\label{op}
\end{eqnarray}
The dimension of the first block of $R_i$ and $D_0$ is $(2L+1)\times (2L+1)$ while the dimension of the second block is $(2L)\times (2L)$. The upper and lower off-diagonal blocks are therefore rectangular matrices with dimensions $(2L+1)\times (2L)$ and $(2L)\times (2L+1)$ respectively. In the above equation the definitions of $R_i^{(l)}$ are the usual ones, i.e ( with $R_{\pm}^{(l)}=R_1^{(l)}\pm i R_2^{(l)}$ and $l=L,L-\frac{1}{2}$ ) 
\begin{eqnarray}
(R_{\pm}^{(l)})_{ll_3\pm 1,ll_3}=\sqrt{(l\mp l_3)(l\pm l_3+1)}~,~(R_3^{(l)})_{ll_3,ll_3}=l_3,
\end{eqnarray}
whereas $V_{\alpha}^{(L,L-\frac{1}{2})}$ and  $V_{\alpha}^{(L-\frac{1}{2},L)}$ are given by 
\begin{eqnarray}
(V_{\pm}^{(L,L-\frac{1}{2})})_{Ll_3\pm \frac{1}{2},L-\frac{1}{2}l_3}=-\frac{1}{2}\sqrt{L\pm l_3+\frac{1}{2}}~,~(V_{\pm}^{(L-\frac{1}{2},L)})_{L-\frac{1}{2}l_3\pm \frac{1}{2},Ll_3}=\mp \frac{1}{2}\sqrt{L\mp l_3}.
\end{eqnarray} 
We will also denote the operators given in (\ref{op}) by ${\Lambda}_i^{(L)}\equiv R_i$, $i=1,2,3$,  ${\Lambda}_{\alpha}^{(L)}$ ( $\equiv$ $V_{\pm}$, $D_{\pm}$ ), $\alpha=4,5,6,7$ and ${\Lambda}_8^{(L)}\equiv D_0$. For $L=\frac{1}{2}$ we get the $3-$dimensional fundamental representation of $OSP(2,2)$ given in (\ref{17}) and (\ref{uuz}).

The above irreducible representation with superspin $L$ is characterized by the value of the $OSP(2,1)$ Casimir operator $K_{2,1}=R_i^2+C_{\alpha \beta}V_{\alpha}V_{\beta}$ which is equal $L(L+\frac{1}{2})$ in this representation, viz 
\begin{eqnarray}
K_{2,1}=R_i^2+C_{\alpha \beta}V_{\alpha}V_{\beta}=L(L+\frac{1}{2}).
\end{eqnarray}
The above operators (\ref{op}) give also a non-typical irreducible representation of $OSP(2,2)$ characterized by the value of the $OSP(2,2)$ Casimir operator 
\begin{eqnarray}
K_{2,2}=R_i^2+{C}_{\alpha \beta}V_{\alpha}V_{\beta}-{C}_{\alpha \beta}D_{\alpha}D_{\beta}-\frac{1}{4}D_0^2=0. 
 \end{eqnarray}
This means in particular two things, $1)$ this representation ( as opposed to typical ones of $OSP(2,2)$ ) is irreducible with respect to the $OSP(2,1)$ subgroup  and $2)$ the $OSP(2,2)$ generators $D_{\alpha}$ and $D_0$ can be realized nonlinearly in terms of the $OSP(2,1)$ generators.

The space ${\cal L}(L,L)$ is isomorphic to the algebra of supermatrices $Mat(2L+1,2L)$. The dimension of the Hilbert space on which this algebra acts is $N=(2L+1)+(2L)=4L+1$. The coordinates operators on the fuzzy supersphere are defined by $\hat{n}_i=2R\hat{\Omega}_i$, $\hat{n}_{4,5}=2R\hat{\Omega}_{4,5}$ where
\begin{eqnarray}
\hat{\Omega}_i=\frac{R_i}{2\sqrt{L(L+\frac{1}{2})}}~,~\hat{\Omega}_4=\frac{V_+}{2\sqrt{L(L+\frac{1}{2})}}~,~\hat{\Omega}_5=\frac{V_-}{\sqrt{2L(L+\frac{1}{2})}}.
\end{eqnarray} 
 The remaining coordinates operators $\hat{n}_{6,7,8}=2R\hat{\Omega}_{6,7,8}$ are similarly defined by
\begin{eqnarray}
\hat{\Omega}_8=\frac{D_0}{2\sqrt{L(L+\frac{1}{2})}}~,~\hat{\Omega}_6=\frac{D_+}{2\sqrt{L(L+\frac{1}{2})}}~,~\hat{\Omega}_7=\frac{D_-}{2\sqrt{L(L+\frac{1}{2})}}.
\end{eqnarray} 
These coordinates operators satisfy the commutation and anticommutation relations ( with $a,b,c=1,...,8$ )
\begin{eqnarray}
[\hat{n}_a,\hat{n}_b\}=\hat{n}_a\hat{n}_b-(-1)^{d_{n_a}d_{n_b}}\hat{n}_b\hat{n}_a=\frac{iR}{\sqrt{L(L+\frac{1}{2})}}f_{abc}\hat{n_c}.
\end{eqnarray}
The definition of the structure constants $f_{abc}$ is obvious from (\ref{834}) and (\ref{87g}). These coordinates operators must also satisfy the constraints 
\begin{eqnarray}
\hat{n}_i^2+C_{\alpha \beta}\hat{n}_{\alpha}\hat{n}_{\beta}=R^2.
\end{eqnarray} 
\begin{eqnarray}
\hat{n}_i^2+\tilde{{C}}_{\alpha \beta}\hat{n}_{\alpha}\hat{n}_{\beta}-\frac{1}{4}\hat{n}_8^2=0. 
\end{eqnarray}
The continuum limit is defined by $L{\longrightarrow}\infty $  in which $\hat{n}_a{\longrightarrow}n_a$ and $\hat{\Omega}_a{\longrightarrow}{\Omega}_a$. 
To see this more explicitly we notice that under the adjoint action of $OSP(2,1)$ the algebra $Mat(2L+1,2L)$ decomposes as 
\begin{eqnarray}
Mat(2L+1,2L)\equiv L\otimes L=0\oplus \frac{1}{2}\oplus 1\oplus ...\oplus 2L-\frac{1}{2}\oplus 2L.
\end{eqnarray}
The dimension of this space is $N^2$ and a generic element  is a polynomial in  $\hat{n}_{i,4,5}$. Recall that $\hat{n}_{6,7,8}$ can be realized nonlinearly in terms of the $\hat{n}_{i,4,5}$. Among these polynomials we can define the matrix superspherical harmonics. A given $N\times N$  supermatrix can be expanded  in terms of these superspherical harmonics. In the continuum limit $Mat(2L+1,L)$ approaches the algebra of superfunctions on the supersphere. In partiuclar the matrix superspherical harmonics go to the ordinary superspherical harmonics which are the eigensuperfunctions of the Casimir operator ${\cal R}_i^2+C_{\alpha \beta}{\cal V}_{\alpha}{\cal V}_{\beta}$ and  ${\cal R}_3$. 

A very important remark is to note that elements of $Mat(2L+1,2L)$ ( in other words superfields ) can be even or odd  if $Mat(2L+1,2L)$ is defined over a graded commutative algebra ${\bf P}$ instead of the field of complex numbers. In this case we will denote this algebra by $Mat(2L+1,2L;P)$. In the fuzzy case we have the definitions ${\cal R}_i\Phi =[R_i,\Phi]$ and ${\cal V}_{\alpha}{\Phi}_{\rm odd} =\{V_{\alpha},{\Phi}_{\rm odd}\}$, ${\cal V}_{\alpha}{\Phi}_{\rm even} =[V_{\alpha},{\Phi}_{\rm even}]$ where $\Phi$ is any element of $Mat(2L+1,2L;P)$, ${\Phi}_{\rm odd}$ is an odd element of $Mat(2L+1,2L;P)$ and ${\Phi}_{\rm even}$ is an even element of $Mat(2L+1,2L;P)$. Strictly speaking the fuzzy supersphere is identified with the even elements of $Mat(2L+1,2L;P)$ while the odd elements will be crucial in constructing gauge theories.
The inner product on $Mat(2L+1,2L;P)$ is defined by
\begin{eqnarray}
({\Phi}_1,{\Phi}_2)\equiv STr{\Phi}_1^{++}{\Phi}_2.
\end{eqnarray}
This  satisfies $STr({\bf 1}_N)=1$ and $STr[X,Y\}=0$.

A general supermatrix ${\Phi}{\in}Mat(2L+1,2L;P)$ and its graded involution ${\Phi}^{++}$ are given by
\begin{eqnarray}
&&\Phi=\bigg(\begin{array}{cc}
                   {\phi}_R& {\psi}_R \\
            {\psi}_L & {\phi}_L  
                 \end{array}\bigg)~,~{\Phi}^{++}=\bigg(\begin{array}{cc}
                   {\phi}_R^{++} & \mp {\psi}_L^{++} \\
            \pm {\psi}_R^{++} & {\phi}_L^{++}
                 \end{array}\bigg).
\end{eqnarray}
${\phi}_R$ and ${\phi}_L$ are $(2L+1)\times (2L+1)$ and $(2L)\times (2L)$ matrices while ${\psi}_R$ and ${\psi}_L$ are respectively $(2L+1)\times (2L)$ and $(2L)\times (2L+1)$ matrices.  In ${\Phi}^{++}$ the upper signs refer to the case when $\Phi$ is  an even superfield ( in which case the off-diagonal blocks are fermionic and the diagonal blocks are bosonic), while the lower  signs refer to the case when $\Phi$ is an odd ( in which case the off-diagonal blocks are bosonic and the diagonal blocks are fermionic ).  We remark that $STr\Phi =Tr_{2L+1}{\phi}_R-(-1)^{|\Phi|}Tr_{2L}{\phi}_L$.

The Laplacian on the fuzzy supersphere is given by
\begin{eqnarray}
{\Delta}={\cal K}_{2,1}-{\cal K}_{2,2}=\frac{1}{4}{\cal D}_0^2+{\cal D}_6{\cal D}_7-{\cal D}_7{\cal D}_6.
\end{eqnarray}
The definition of ${\cal D}_{0,6,7}$ are obvious by analogy with ${\cal R}_i$ and ${\cal V}_{4,5}$ given above. The fuzzy supersphere of size $N=4L+1$ is by definition the spectral triple consisting of $1)$ the algebra of supermatrices $Mat(2L+1,2L)$ together with $2)$ the  representation space of the superspin $L$ of $OSP(2,1)$ with inner product given by the supertrace $STr$ and graded involution given by $++$ and  $3)$ the Laplacian ${\Delta}$ which is  the most important ingredient. The Laplacian  fixes the metric aspects of the space uniquely while the algebra alone will only give  toplogy.
\section{Gauge theory}
\subsection{Klimcik differential complex}
The  Laplacian on the fuzzy supersphere depends only on the $OSP(2,2)$ generators $D_{\pm,0}$ in the adjoint representation. This means in particular that the $OSP(2,2)$  generators in the directions $D_{\pm}$ are the supersymmetric covariant derivatives on the fuzzy supersphere while the $OSP(2,2)$ generators $V_{\pm}$ are  the supersymmetry generators. A gauge field on the supersphere is a  superspin $1/2$ multiplet  composed of $3$ superfields $A_{\pm}$ and $W$ in the directions $D{\pm}$ and $D_0$ resepctively. These superfields $A_{\pm}$ and $W$ transform under $OSP(2,1)$ in the same way as $D_{\pm}$ and $D_0$. The supercovariant derivatives ( as opposed to the covariant derivatives in the non-supersymmetric case ) are thus
\begin{eqnarray}
X_{\pm}=D_{\pm}+A_{\pm}~,~X_0=D_0+W.
\end{eqnarray} 
In order to construct   gauge theory on the fuzzy  supersphere we must in fact start from an $OSP(2,2)$ supervector. Thus we need to add a superspin $1$ multiplet composed of $5$ more superfields $C_i$ and $B_{\pm}$ ( which transform under $OSP(2,1)$ in the same way as $V_{\pm}$ and $R_i$ ) with  supercovariant derivatives 
\begin{eqnarray}
Y_i=R_i+C_i~,~Z_{\pm}=V_{\pm}+B_{\pm}.
\end{eqnarray} 
In the following we will construct explicitly the  action principle of the $OSP(2,2)$ vector gauge superfield $(A_{\pm},W,C_i,B_{\pm})$. In the case of the fuzzy supersphere this action principle will be a supermatrix model. We will also need to write down constraints which must be satisfied by these superfields in order to have the correct number of degrees of freedom on the fuzzy supersphere.

The differential complex over the fuzzy supersphere is defined by
\begin{eqnarray}
{\Psi}_N={\oplus}_{j=0}^{3}{\Psi}_N^j
\end{eqnarray}
The elements of ${\Psi}_N^j$ are the $j-$forms. We have the following identifications
\begin{eqnarray}
&&{\Psi}_N^0={\Psi}_N^3=Mat(2L+1,2L)~,~{\Psi}_N^1={\Psi}_N^2={\otimes}_{i=1}^8Mat(2L+1,2L)_i.
\end{eqnarray}
´We must clearly have $Mat(2L+1,2L)_i=Mat(2L+1,2L)$. A zero-form is thus an element $\Phi$ of the algebra ${\Psi}_N^0=Mat(2L+1,2L)$ while 
a one-form  is an element of ${\Psi}_N^1$ of the form
\begin{eqnarray}
A=
(A_{\pm},W,C_i,B_{\pm}).
\end{eqnarray}
The $5$ superfields  $C_{i}$, $B_{\pm}$ transform as a superspin $1$ multiplet under $OSP(2,1)$ while the remaining $3$ superfields $A_{\pm}$ and $W$ will transform as a superspin $1/2$ under $OSP(2,1)$.  All these superfields are elements of the algebra $Mat(2L+1,2L)$. We can also write zero-forms and one-forms as  $3N\times 3N$ supermatrices of the form
\begin{eqnarray}
{M}_1
&=& r_+\otimes C_-+r_-\otimes C_++2r_3\otimes C_3 +2v_+\otimes B_{-} -2v_-\otimes B_+ -2d_+\otimes A_{-}+ 2d_-\otimes A_+\nonumber\\
&-&\frac{1}{2}d_0\otimes W.\label{sur116}\nonumber\\
M_0&=&{\bf 1}_3\otimes \Phi.
\end{eqnarray}
In above ${\Lambda}_i^{(\frac{1}{2})}\equiv r_i$, $i=1,2,3(\pm ,3)$,  ${\Lambda}_{\alpha}^{(\frac{1}{2})}=v_{\alpha}$, $\alpha=4(+),5(-)$,   ${\Lambda}_{\alpha}^{(\frac{1}{2})}=d_{\alpha}$, $\alpha=6(+),7(-)$ and ${\Lambda}_8^{(\frac{1}{2})}\equiv d_0$ are the supermatrices of the $3-$dimensional superspin $1/2$ fundamental representation of $OSP(2,2)$ corresponding to the generators $R_i$, $V_{\pm}$, $D_{\pm}$ and $D_0$ respectively. We will also use the notation ${\Lambda}_i\equiv R_i$, $i=1,2,3(\pm,3)$,  ${\Lambda}_{\alpha}=V_{\alpha}$, $\alpha=4(+),5(-)$,   ${\Lambda}_{\alpha}=D_{\alpha}$, $\alpha=6(+),7(-)$ and ${\Lambda}_8\equiv D_0$.

Similarly two-forms and three-forms are given by  $a=
(a_{\pm},w,c_i,b_{\pm}){\in}{\Psi}_N^2={\Psi}_N^1$ and $\phi {\in}{\Psi}_N^3={\Psi}_N^0$. We write the corresponding $3N\times 3N$ supermatrices as 
\begin{eqnarray}
{M}_2
&=& r_+\otimes c_-+r_-\otimes c_++2r_3\otimes c_3 +2v_+\otimes b_{-} -2v_-\otimes b_+ -2d_+\otimes a_{-}+ 2d_-\otimes a_+-\frac{1}{2}d_0\otimes w.\label{sur117}\nonumber\\
M_3&=&{\bf 1}_3\otimes \phi.
\end{eqnarray}
Let us introduce the quadratic Casimirs 
\begin{eqnarray}
C_G&=&r_+\otimes R_-+r_-\otimes R_++2r_3\otimes R_3+2v_+\otimes V_--2v_-\otimes V_+-2d_+\otimes D_-+2d_-\otimes D_+\nonumber\\
&-&\frac{1}{2}d_0\otimes D_0
\nonumber\\
C_H&=&r_+\otimes R_-+r_-\otimes R_++2r_3\otimes R_3+2v_+\otimes V_--2v_-\otimes V_+\nonumber\\
C&=&C_G-C_H=-2d_+\otimes D_-+2d_-\otimes D_+-\frac{1}{2}d_0\otimes D_0.
\end{eqnarray}
\paragraph{The exterior derivative}
We introduce a coboundary  operator  ${\delta}:{\Psi}_N^i{\longrightarrow}{\Psi}_N^{i+1}$ defined on $0-$forms by 
\begin{eqnarray}
{\delta}\Phi&=&\bigg(r_+\otimes {\cal R}_-+r_-\otimes {\cal R}_++2r_3\otimes {\cal R}_3+2v_+\otimes {\cal V}_--2v_-\otimes {\cal V}_+\nonumber\\
&-&2d_+\otimes {\cal D}_-+2d_-\otimes {\cal D}_+-\frac{1}{2}d_0\otimes {\cal D}_0\bigg)\Phi.
\end{eqnarray}
The exterior derivative on one-forms is on the other hand given by 
\begin{eqnarray}
&&{\delta}M_1={\delta}^GM_1-{\delta}^HM_1^{H}.
\end{eqnarray}
$M_1^{H}$ is the orthogonal projection of $M_1$ from $G\otimes Mat(2L+1,2L)$ into $H\otimes Mat(2L+1,2L)$ where $G=osp(2,2)$ and $H=osp(2,1)$. In other words
\begin{eqnarray}
M_1^{H}
&=& r_+\otimes C_-+r_-\otimes C_++2r_3\otimes C_3 +2v_+\otimes B_{-} -2v_-\otimes B_+.
\end{eqnarray}
The exterior derivatives ${\delta}^G$ and ${\delta}^H$ are defined by ( with the notation $M_1\equiv h_A\otimes C_A$ and $adO\equiv {\cal O}=[O,.]$)

\begin{eqnarray}
&&{\delta}^GM_1=2(-1)^{h\Lambda}\hat{\eta}_{ab}ad{\Lambda}_{a}^{(\frac{1}{2})}h_A\otimes ad{\Lambda}_{b}C_A+\frac{1}{2}d_GM_1\nonumber\\
&&{\delta}^HM_1=2(-1)^{h\Lambda}\bar{\eta}_{ab}ad{\Lambda}_{a}^{(\frac{1}{2})}h_A\otimes ad{\Lambda}_{b}C_A+\frac{1}{2}d_HM_1.
\end{eqnarray}
$\hat{\eta}$, $\bar{\eta}$ stand for the block diagonal matrices $\hat{\eta}=2({\bf 1}_3,\tilde{C},-1/4)$, $\bar{\eta}=2({\bf 1}_3,{C})$ and $a,b =1,...,8$ in  ${\delta}^GM_1$ and $a,b=1,...,5$ in ${\delta}^HM_1$. The Dynkin numbers $d_G,d_H$ are defined by 
\begin{eqnarray}
STrXY=\frac{4L^2}{d_G}STr_G(X_{1}Y_{1})~,~STrXY=\frac{4L^2}{d_H}STr_H(X_{1}Y_{1}).
\end{eqnarray}
where $STr_G$, $STr_H$ are the supertraces in the adjoint representations of $G$ and $H$ respectively. We choose $X=Y=R_i$ so $XY=R_i^2=diag((R_i^{(L)})^2,(R_i^{(L-\frac{1}{2})})^2)$ for $H$ and $XY=R_i^2=diag((R_i^{(L)})^2,(R_i^{(L-\frac{1}{2})})^2;(R_i^{(L-\frac{1}{2})})^2,(R_i^{(L-1)})^2)$ for $G$. The supermatrices $X_{1}$ and $Y_{1}$ correspond to the representation $L=1$. Using the property $STr\Phi =Tr_{2L+1}{\phi}_R-Tr_{2L}{\phi}_L$ of $STr$ we compute for $G$ that $STrR_i^2=6L^2$ and hence $d_G=6$ while for $H$ we compute $STrR_i^2=3L(L+\frac{1}{2})$ and hence $d_H=6+3/L$.

We have explicitly
\begin{eqnarray}
{\delta}M_1={\delta}^GM_1^{\perp}+({\delta}^G-{\delta}^H)M_1^H~,~M_1^{\perp}=M_1-M_1^H.
\end{eqnarray}
We can immediately compute
\begin{eqnarray}
2{\delta}^G(d_{\pm}\otimes A_{\mp})&=&4d_{\mp}\otimes {\cal R}_{\pm}A_{\mp}{\pm}4d_{\pm}\otimes {\cal R}_3A_{\mp}+2d_0{\otimes}{\cal V}_{\pm}A_{\mp}-2v_{\pm}\otimes {\cal D}_0A_{\mp}\nonumber\\
&-&4r_{\pm}\otimes {\cal D}_{\mp}A_{\mp}{\mp}4r_3{\otimes}{\cal D}_{\pm}A_{\mp}+d_Gd_{\pm}\otimes A_{\mp},
\end{eqnarray}
and
\begin{eqnarray}
\frac{1}{2}{\delta}^G(d_0\otimes W)&=&-2d_{+}\otimes {\cal V}_-W+ 2d_-\otimes {\cal V}_+W +2v_+\otimes {\cal D}_{-}W-2v_{-}\otimes {\cal D}_{+}W\nonumber\\
&+&\frac{1}{4}d_Gd_0\otimes W.
\end{eqnarray}
Hence
\begin{eqnarray}
{\delta}^GM_1^{\perp}&=&-4r_{-}\otimes {\cal D}_+A_+ +4r_{+}\otimes {\cal D}_{-}A_{-}+4r_{3}\otimes ({\cal D}_+A_- +{\cal D}_-A_+)\nonumber\\
&-&2v_{-}\otimes ({\cal D}_0A_+-{\cal D}_+W)+2v_+\otimes ({\cal D}_0A_{-}-{\cal D}_{-}W)\nonumber\\
&-&d_{-}\otimes (4{\cal R}_+A_- +4{\cal R}_3A_++2{\cal V}_+W -d_G A_+)\nonumber\\
&+&d_{+}\otimes (4{\cal R}_-A_+-4{\cal R}_3A_- +2{\cal V}_- W-d_GA_-)\nonumber\\
&+&d_0\otimes (-2{\cal V}_+A_- +2{\cal V}_-A_+-\frac{1}{4}d_GW).
\end{eqnarray}
Furthermore ( with the notation $M_1^H=h_A\otimes C_A$ ) 
\begin{eqnarray}
({\delta}^G-{\delta}^H)M_1^H&=&(-1)^{h_A}(-4add_+\otimes {\cal D}_-+4add_-\otimes {\cal D}_+-add_0\otimes {\cal D}_0)(h_A\otimes C_A)\nonumber\\
&+&\frac{1}{2}(d_G-d_H)M_1^H.
\end{eqnarray}
We compute
\begin{eqnarray}
2(-4add_+\otimes {\cal D}_-+4add_-\otimes {\cal D}_+-add_0\otimes {\cal D}_0)(r_i\otimes C_i)=4d_+\otimes ({\cal D}_-C_3-{\cal D}_+C_-)\nonumber\\
+4d_-\otimes ({\cal D}_+C_3+{\cal D}_-C_+),
\end{eqnarray}
and
\begin{eqnarray}
-2(-4add_+\otimes {\cal D}_-+4add_-\otimes {\cal D}_++add_0\otimes {\cal D}_0)(v_+\otimes B_--v_-\otimes B_+)&=&\nonumber\\
-2d_+\otimes {\cal D}_0B_-+2d_-\otimes {\cal D}_0B_++2d_0\otimes ({\cal D}_+B_--{\cal D}_-B_+).
\end{eqnarray}
Thus
\begin{eqnarray}
({\delta}^G-{\delta}^H)M_1^H&=&\frac{1}{2}(d_G-d_H)(r_+\otimes C_-+r_-\otimes C_++2r_3\otimes C_3 +2v_+\otimes B_{-} -2v_-\otimes B_+)\nonumber\\
&+&2d_+\otimes (-{\cal D}_0B_-+2{\cal D}_-C_3-2{\cal D}_+C_-)-2d_-\otimes (-{\cal D}_0B_+-2{\cal D}_+C_3-2{\cal D}_-C_+)\nonumber\\
&+&2d_0\otimes ({\cal D}_+B_--{\cal D}_-B_+).
\end{eqnarray}
The final result is ( with $d_G=4$, $d_H=6$ )
\begin{eqnarray}
&&{\delta}A_{\pm}=2A_{\pm}+2{\cal D}_{\mp}C_{\pm}-2{\cal R}_{\pm}A_{\mp}{\pm}2{\cal D}_{\pm}C_3{\mp}2{\cal R}_3A_{\pm}+{\cal D}_0B_{\pm}-{\cal V}_{\pm}W\nonumber\\
&&{\delta}W=2W+4{\cal V}_+A_--4{\cal V}_-A_++4{\cal D}_-B_+-4{\cal D}_+B_-\nonumber\\
&&{\delta}B_{\pm}=-B_{\pm}+{\cal D}_0A_{\pm}-{\cal D}_{\pm}W\nonumber\\
&&{\delta}C_3=-C_3+2{\cal D}_+A_-+2{\cal D}_-A_+\nonumber\\
&&{\delta}C_{\pm}=-C_{\pm}{\mp}4{\cal D}_{\pm}A_{\pm}.
\end{eqnarray}
Lastly the action of the coboundary operator on two-forms a three-forms is given by the obvious definitions ( with $M_2=h_A\otimes c_A$ )
\begin{eqnarray}
{\delta}M_2&=&-\frac{1}{2}{\bf 1}_3\otimes {\cal H}_Ac_A\nonumber\\
&=&{\cal D}_+a_--{\cal D}_-a_++\frac{1}{4}{\cal D}_0w-{\cal V}_+b_-+{\cal V}_-b_+-\frac{1}{2}{\cal R}_+c_--\frac{1}{2}{\cal R}_-c_+-{\cal R}_3c_3.
\end{eqnarray}
\begin{eqnarray}
{\delta}M_3={\delta}({\bf 1}_3\otimes \phi)=0.
\end{eqnarray}

\paragraph{The product $*$}

The associative product $*$ between the forms is a map $*:{\Psi}_N^i\otimes {\Psi}_N^j{\longrightarrow}{\Psi}_N^{i+j}$ defined for $i=1$ by ( with $h_A\otimes X_A$ standing for one-forms, two-forms and three-forms )
\begin{eqnarray}
({\bf 1}_3\otimes \Phi )*(h_A\otimes X_A)=h_A\otimes \Phi X_A.
\end{eqnarray}
For $i=2$ we have
\begin{eqnarray}
&&(h_A\otimes C_A)*({\bf 1}_3\otimes \Phi)=h_A\otimes C_A\Phi~,~(h_A\otimes C_A)*({\bf 1}_3\otimes \phi)=0,
\end{eqnarray}
and
\begin{eqnarray}
(h_A\otimes C_A)*(h_A^{'}\otimes C_A^{'})=(h_A\otimes C_A)*_G(h_A^{'}\otimes C_A^{'})-(h_A\otimes C_A)*_H(h_A^{'}\otimes C_A^{'}),
\end{eqnarray}
where
\begin{eqnarray}
(h_A\otimes C_A)*_{G,H}(h_A^{'}\otimes C_A^{'})=2(-1)^{Ch^{'}}ad(h_A)h_{B}^{'}\otimes C_AC_{B}^{'}\nonumber\\
\end{eqnarray}
The indices $A$ and $B$  run over the superalgebra $H$ for $*_H$ whereas for $*_G$ they run over the superalgebra $G$. Explicitly we have
\begin{eqnarray}
(h_A\otimes C_A)*(h_A^{'}\otimes C_A^{'})&=&2\big(r_i\otimes C_i+v_+\otimes B_--v_-\otimes B_+\big)*_G\big(-2d_+\otimes A_-^{'}+2d_-\otimes A_+^{'}\nonumber\\
&-&\frac{1}{2}d_0\otimes W^{'}\big)+\big(-2d_+\otimes A_-+2d_-\otimes A_+-\frac{1}{2}d_0\otimes W\big)*_G(h_A^{'}\otimes C_A^{'}).\nonumber\\
\end{eqnarray}
The first line is computed to be given by
\begin{eqnarray}
{\rm First}~{\rm line}&=&2d_+\otimes ( B_-W^{'}+2C_-A_+^{'}-2C_3A_-^{'})-2d_{-}\otimes (B_+W^{'}+2C_+A_-^{'}+2C_3A_+^{'})\nonumber\\
&+&2d_0\otimes (B_-A_{+}^{'}-B_+A_{-}^{'}).
\end{eqnarray}
The second line is computed to be given by
\begin{eqnarray}
{\rm Second}~{\rm line}&=&2d_+\otimes (- WB_-^{'}+2A_-C_3^{'}-2A_+C_-^{'})+2d_{-}\otimes (WB_+^{'}+2A_+C_3^{'}+2A_-C_+^{'})\nonumber\\
&-&2d_0\otimes (A_-B_{+}^{'}-A_+B_{-}^{'})+4r_+\otimes A_-A_-^{'}-4r_-\otimes A_+A_+^{'}+4r_3\otimes (A_-A_+^{'}+A_+A_-^{'})\nonumber\\
&-&2v_+\otimes (A_-W^{'}-WA_-^{'})+2v_-\otimes (A_+W^{'}-WA_+^{'}).
\end{eqnarray}
Thus we obtain
\begin{eqnarray}
&&A_{\pm}*A_{\pm}^{'}={\pm}2A_{\pm}C_3^{'}{\mp}2C_3A_{\pm}^{'}+WB_{\pm}^{'}-B_{\pm}W^{'}+2A_{\mp}C_{\pm}^{'}-2C_{\pm}A_{\mp}^{'}\nonumber\\
&&W*W^{'}=4A_-B_+^{'}+4B_+A_-^{'}-4A_+B_-^{'}-4B_-A_+^{'}\nonumber\\
&&C_{\pm}*C_{\pm}^{'}={\mp}4A_{\pm}A_{\pm}^{'}\nonumber\\
&&C_3*C_3^{'}=2A_-A_+^{'}+2A_+A_-^{'}\nonumber\\
&&B_{\pm}*B_{\pm}=WA_{\pm}^{'}-A_{\pm}W^{'}.
\end{eqnarray}
Also for $i=2$ we have
\begin{eqnarray}
&&(h_A\otimes C_A)*(h_{A}^{'}\otimes c_{A}^{'})=-\frac{1}{2}(-1)^{h^{'}C}STr(h_Ah_{B}^{'})\otimes C_Ac_{B}^{'}.
\end{eqnarray}
By using the identities $STrr_ir_j=\frac{1}{2}{\delta}_{ij}$, $STrv_{\pm}v_{\mp}={\mp}\frac{1}{2}$, $STrd_{\pm}d_{\mp}={\pm}\frac{1}{2}$ and $STrd_0^2=-2$ ( all other supertraces are zero ) we obtain immdediately the results
\begin{eqnarray}
(h_A\otimes C_A)*(h_{A}^{'}\otimes c_{A}^{'})&=&-\frac{1}{2}C_+c_-^{'}-\frac{1}{2}C_-c_+^{'}-C_3c_3^{'}+A_+a_-^{'}-A_-a_+^{'}+\frac{1}{4}Ww^{'}\nonumber\\
&+&B_-b_+^{'}-B_+b_-^{'}.
\end{eqnarray}
For $i=3$ we have the two non-vanishing products $(h_A\otimes c_A)*({\bf 1}_3\otimes \Phi)$ and $(h_A\otimes c_A)*(h_A^{'}\otimes C_A^{'})$ with obvious 
definitions by analogy with the products $({\bf 1}_3\otimes \Phi)*(h_A\otimes c_A)$ and $(h_A^{'}\otimes C_A^{'})*(h_A\otimes c_A)$. In particular the product of two-forms with one-forms is given as above with reversed order of small and capital letters. For $i=4$ we have one non-zero product given by $({\bf 1}_3\otimes \phi)*({\bf 1}_3\otimes \Phi)$ while the rest are zero.

This coboundary operator is nilpotent, i.e it satisfies ${\delta}^2=0$.
The product $*$ is compatible with ${\delta}$ so that the Leibniz rule is respected. Thus we must have ${\delta}(X^i*Y^j)={\delta}X^i*Y^j+(-1)^{i}X^i*{\delta}Y^j$.
\subsection{Gauge action}
We consider one-forms $A=(A_{\pm},W,C_i,B_{\pm})$ satisfying the reality condition $A^{++}=A$. Thus we must have $A_{\pm}^{++}=\pm A_{\mp}$, $W^{++}=W$, $C_i^{++}=C_i$ and $B_{\pm}^{++}={\mp}B_{\mp}$. We define the curvature by
\begin{eqnarray}
F={\delta}A+A*A=(F_{\pm},f,c_i,b_{\pm}).
\end{eqnarray}
We can immediately compute
\begin{eqnarray}
&&F_{\pm}=2[X_{\mp},Y_{\pm}]{\pm}2[X_{\pm},Y_3]+[X_0,Z_{\pm}]+2X_{\pm}\nonumber\\
&&f=4\{Z_+,X_-\}-4\{Z_-,X_+\}+2X_0\nonumber\\
&&c_{\pm}={\mp}2\{X_{\pm},X_{\pm}\}-Y_{\pm}\nonumber\\
&&c_3=2\{X_+,X_-\}-Y_3\nonumber\\
&&b_{\pm}=[X_0,X_{\pm}]-Z_{\pm}.
\end{eqnarray}
Recall that $X_{\pm}=D_{\pm}+A_{\pm}$, $X_0=D_0+W$, $Y_i=R_i+C_i$ and $Z_{\pm}=V_{\pm}+B_{\pm}$. We define the supersymmetric noncommutative $U(1)$  gauge action by 
\begin{eqnarray}
S_L[A]&=&{\alpha}Str\triangleleft F*F+\beta STr(A*{\delta}A+\frac{2}{3}A*A*A)\nonumber\\
&=&{\alpha}Str\triangleleft F*F+\beta STr(A*F-\frac{1}{3}A*A*A).
\end{eqnarray}
The first term is similar to the usual Yang-Mills action whereas the second term is a ( real-valued ) Chern-Simons-like contribution. $\alpha$ and $\beta$ are two real parameters. The Hodge triangle $\triangleleft $ is defined as the identity map between ${\Psi}_N^1$ and ${\Psi}_N^2$ and thus $\triangleleft F$ should be considered as a one-form. Explicitly we have
\begin{eqnarray}
&&\triangleleft F*F=b_-b_+-b_+b_--c_i^2+F_+F_--F_-F_++\frac{1}{4}f^2\nonumber\\
&&A*F=B_-b_+-B_+b_--C_ic_i+A_+F_--A_-F_++\frac{1}{4}Wf,
\end{eqnarray}
and
\begin{eqnarray}
A*A*A&=&B_-[W,A_+]-B_+[W,A_-]-2C_+A_-A_-+2C_-A_+A_+-2C_3\{A_+,A_-\}\nonumber\\
&+&A_+\big(2[C_3,A_-]-2[C_-,A_+]+[W,B_-]\big)-A_-\big(-2[C_3,A_+]-2[C_+,A_-]+[W,B_+]\big)\nonumber\\
&+&W\{A_-,B_+\}-W\{A_+,B_-\}.
\end{eqnarray}
We need first to show gauge invariance of the above action. The invariance of the Yang-Mills term is obvious whereas the invariance of the Chern-Simons-like term requires some work in order to be established. Towards this end we will need to rewrite the Chern-Simons-like term in a completely covariant fashion. 

First by thinking about $C_G$ as a one-form we can show after a long calculation that we must have
\begin{eqnarray}
STrC_G*F&=&STr\bigg(-2D_+A_-+2D_-A_++V_-B_+-V_+B_--R_iC_i-\frac{1}{2}D_0W\bigg)\nonumber\\
&+&\frac{1}{2} STrW\bigg(\{D_-,B_+\}-\{D_+,B_-\}-\{V_-,A_+\}+\{V_+,A_-\}\bigg)\nonumber\\
&-&STrA_+\bigg([D_-,C_3]-[D_+,C_-]-\frac{1}{2}[D_0,B_-]+[R_-,A_+]-[R_3,A_-]-\frac{1}{2}[W,V_-]\bigg)\nonumber\\
&-&STrA_-\bigg([D_-,C_+]+[D_+,C_3]+\frac{1}{2}[D_0,B_+]-[R_+,A_-]-[R_3,A_+]+\frac{1}{2}[W,V_+]\bigg)\nonumber\\
&-&STrC_+\{D_-,A_-\}+STrC_-\{D_+,A_+\}-\frac{1}{2}STrB_+\big([D_0,A_-]-[D_-,W]\big)\nonumber\\
&+&\frac{1}{2}STrB_-\big([D_0,A_+]-[D_+,W]\big)-STrC_3\bigg(\{D_-,A_+\}+\{D_+,A_-\}\bigg).
\end{eqnarray}
In above we have used the results
\begin{eqnarray}
STrD_{\mp}F_{\pm}&=&STr\bigg({\pm}R_{\mp}C_{\pm}+2D_{\pm}A_{\mp}{\pm}R_3C_3+D_{\mp}A_{\pm}-V_{\mp}B_{\pm}{\pm}\frac{1}{4}D_0W\nonumber\\
&+&2A_{\mp}[D_{\mp},C_{\pm}]{\pm}2A_{\pm}[D_{\mp},C_3]-W\{D_{\mp},B_{\pm}\}\bigg)\nonumber\\
STrV_{\mp}b_{\pm}
&=&STr\bigg(-D_{\mp}A_{\pm}{\mp}\frac{1}{4}D_0W-V_{\mp}B_{\pm}-W\{V_{\mp},A_{\pm}\}\bigg)\nonumber\\
\frac{1}{4}STrD_0f&=&STr\bigg(D_+A_--D_-A_++V_-B_+-V_+B_-+\frac{1}{2}D_0W-A_-[D_0,B_+]+A_+[D_0,B_-]\bigg)\nonumber\\
STrR_{\mp}c_{\pm}&=&STr\bigg(\mp  4D_{\mp}A_{\pm}-R_{\mp}C_{\pm}{\pm}2A_{\pm}[R_{\mp},A_{\pm}]\bigg)\nonumber\\
STrR_3c_3&=&STr\bigg(D_+A_--D_-A_+-R_3C_3-A_-[R_3,A_+]-A_+[R_3,A_-]\bigg).
\end{eqnarray}
More calculation yields
\begin{eqnarray}
STrC_G*F&=&STr\bigg(-D_+A_-+D_-A_+-V_-B_++V_+B_--R_iC_i+\frac{1}{2}D_0W\bigg)\nonumber\\
&+&\frac{1}{2} STrW\bigg(\frac{1}{4}f-\frac{1}{2}W-\{A_-,B_+\}+\{B_-,A_+\}\bigg)\nonumber\\
&-&STrA_+\bigg(-\frac{1}{2}F_-+A_--[A_-,C_3]+[A_+,C_-]+\frac{1}{2}[W,B_-]\bigg)\nonumber\\
&-&STrA_-\bigg(\frac{1}{2}F_+-A_+-[A_-,C_+]-[A_+,C_3]-\frac{1}{2}[W,B_+]\bigg)\nonumber\\
&-&STrC_+\big(\frac{1}{4}c_{-}+\frac{1}{4}C_{-}-\frac{1}{2}\{A_-,A_-\}\big)-STrC_-\big(\frac{1}{4}c_++\frac{1}{4}C_++\frac{1}{2}\{A_+,A_+\}\big)\nonumber\\
&-&\frac{1}{2}STrB_+\big(b_-+B_--[W,A_-]\big)+\frac{1}{2}STrB_-\big(b_++B_+-[W,A_+]\big)\nonumber\\
&-&STrC_3\bigg(\frac{1}{2}c_3+\frac{1}{2}C_3-\{A_-,A_+\}\bigg).
\end{eqnarray}
Equivalently
\begin{eqnarray}
STrC_G*F&=&\frac{1}{2}STrA*F-\frac{1}{2}STrA*A*A-\frac{1}{2}STr\bigg((C_H+A_H)*\triangleleft (C_H+A_H)-C_H*\triangleleft C_H\bigg)\nonumber\\
&-&STr\bigg((C_{\perp}+A_{\perp})*\triangleleft (C_{\perp}+A_{\perp})-C_{\perp}*\triangleleft C_{\perp}\bigg).
\end{eqnarray}
In above $C_{\perp}=C=C_G-C_H$, $A_H$ is the projection of $A$ in the directions along the generators of  $H=osp(2,1)$ and $A_{\perp}$ is the corresponding orthogonal part. Explicitly we have
\begin{eqnarray}
&&STr(C_H+A_H)*\triangleleft (C_H+A_H)
=STr\big(-Y_i^2-Z_+Z_-+Z_-Z_+\big)\nonumber\\
&&STr(C_{\perp}+A_{\perp})*\triangleleft (C_{\perp}+A_{\perp})
=STr\big(\frac{1}{4}X_0^2+X_+X_--X_-X_+\big).
\end{eqnarray}
The supersymmetric noncommutative $U(1)$  gauge action becomes
\begin{eqnarray}
S_L[A]
&=&{\alpha}Str\triangleleft F*F+\frac{\beta}{3} STr\bigg(2(C_H+A_H)*F+2(C_{\perp}+A_{\perp})*F+(C_H+A_H)*\triangleleft (C_H+A_H)\nonumber\\
&+&2(C_{\perp}+A_{\perp})*\triangleleft (C_{\perp}+A_{\perp})-C_H*\triangleleft C_H-2C_{\perp}*\triangleleft C_{\perp}\bigg).\label{78}
\end{eqnarray}
This establishes gauge invariance of the system under
\begin{eqnarray}
C_G+A{\longrightarrow}U*(C_G+A)*U^{++}.
\end{eqnarray}
$U$ is a zero-form with $U^{++}=U^+$ and hence this transformation law means
\begin{eqnarray}
X_{\pm,0}{\longrightarrow}UX_{\pm,0}U^+~,~Y_i{\longrightarrow}UY_iU^+~,~Z_{\pm}{\longrightarrow}UZ_{\pm}U^+.
\end{eqnarray}
The next step is to notice that the system as it stands contains too many degrees of freedom and hence we must impose some extra constraints in order to reduce the number of independent components of $A$ and $F$ from $8$ to $2$ since we are in two dimensions. We impose
\begin{eqnarray}
({\delta}A+A*A)_H=0~,~{\Leftrightarrow}~b_+=b_-=c_+=c_-=c_3=0,\label{158}
\end{eqnarray}
and
\begin{eqnarray}
(C_{\perp}+A_{\perp})*\triangleleft (C_{\perp}+A_{\perp})-C_{\perp}*\triangleleft C_{\perp}=0.\label{159}
\end{eqnarray}
Both constraints are obviously gauge covariant. 

These two constraints as well as the action (\ref{78}) are  invariant under all $OSP(2,1)$ supersymmetry transformations. Indeed the two quantities $(C_H+A_H)*\triangleleft (C_H+A_H)$ and $(C_{\perp}+A_{\perp})*\triangleleft (C_{\perp}+A_{\perp})$ are separately invariant under $OSP(2,1)$ which is the reason behind the invariance of the second constraint and the $4$th and $5$th terms of the action under $OSP(2,1)$. Furthermore a generic one-form and a generic two-form will always  decompose under $OSP(2,1)$ into a direcrt sum of a superpin $1/2$ multiplet  and a superspin $1$ multiplet. For example for the one-form $A$ and for the two-form $F$ the components $A_+,A_-,W$ and $F_+,F_-,f$ form    $OSP(2,1)$ multiplets with superspin $1/2$ while  the other five components $B_+,B_-,C_i$ of $A$ and $b_+,b_-,c_i$ of $F$ form multiplets with superspin $1$. This is the reason why the first constarint is $OSP(2,1)$ invariant. The invariance of the rest of the action under $OSP(2,1)$  is ovbious since it is covariant under full $OSP(2,2)$.
\subsection{The continuum limit}
The constraint (\ref{159}) reads explicitly
\begin{eqnarray}
[D_+,A_-]-[D_-,A_+]+\frac{1}{4}\{D_0,W\}+[A_+,A_-]+\frac{1}{4}W^2=0.
\end{eqnarray}
In the continuum limit this becomes
\begin{eqnarray}
n_6A_--n_7A_++\frac{1}{4}n_8W=0.\label{lkm}
\end{eqnarray}
After some calculation we get the solution ( by using ${\omega}_6=\frac{1}{2}(\bar{z}_1\theta-z_2\bar{\theta})$, ${\omega}_7=\frac{1}{2}(z_1\bar{\theta}+\bar{z}_2\theta)$, ${\omega}_8=2-\bar{z}z$ and $\bar{z}z+\bar{\theta}\theta =1$ )
\begin{eqnarray}
W=\bar{z}z(\bar{A}\frac{\bar{\theta}}{z_2^+}-A\frac{\theta}{z_2})=\bar{A}\frac{\bar{\theta}}{z_2^+}-A\frac{\theta}{z_2},
\end{eqnarray}
where
\begin{eqnarray}
A_+=\frac{1}{2}(A-\frac{z_1^+}{z_2^+}\bar{A})~,~A_-=-\frac{1}{2}(\bar{A}+\frac{z_1}{z_2}{A}).
\end{eqnarray}
The constraint (\ref{158}) leads to the equations
\begin{eqnarray}
&&B_{\pm}=[D_0,A_{\pm}]-[D_{\pm},W]\nonumber\\
&&C_{\pm}=\mp 4\{D_{\pm},A_{\pm}\}\nonumber\\
&&C_3=2\{D_+,A_-\}+2\{D_-,A_+\}.
\end{eqnarray}
We need now to compute the action ( with $L{\longrightarrow}\infty$ )
\begin{eqnarray}
S_L[A]=\alpha STr(F_+F_--F_-F_++\frac{1}{4}f^2)+\beta STr(A_+F_--A_-F_++\frac{1}{4}Wf).
\end{eqnarray}
Explicitly we have
\begin{eqnarray}
F_{\pm}&=&[D_0,[D_0,A_{\pm}]]+2A_{\pm}-[D_0,[D_{\pm},W]]-[V_{\pm},W]{\mp}12[D_{\mp},\{D_{\pm},A_{\pm}\}]{\pm}12[D_{\pm}^2,A_{\mp}]\nonumber\\
f&=&2W+4\{D_+,[D_-,W]\}-4\{D_-,[D_+,W]\}+4\{V_+,A_-\}-4\{D_+,[D_0,A_-]\}\nonumber\\
&-&4\{V_-,A_+\}+4\{D_-,[D_0,A_+]\}.
\end{eqnarray}
Because of the constraint (\ref{lkm}) we have only two independent superfields. We will work in the local coordinates $t=z_1/z_2$, $\bar{t}=z_1^+/z_2^+$, $b=-\theta/z_2$ and $\bar{b}=-\bar{\theta}/z_2^+$. We introduce the parametrization
\begin{eqnarray}
A_+=\frac{1}{2}(A-\bar{t}\bar{A})~,~A_-=-\frac{1}{2}(\bar{A}+tA)~,~W=\bar{b}\bar{A}-bA.
\end{eqnarray}
In terms of $t$, $\bar{t}$ and $b$,$\bar{b}$ the supersymmetryic covariant derivatives $D$,$\bar{D}$ and the supersymmetric charges $Q$,$\bar{Q}$ are given respectively by 
\begin{eqnarray}
D={\partial}_b+b{\partial}_t~,~\bar{D}={\partial}_{\bar{b}}+\bar{b}{\partial}_{\bar{t}},
\end{eqnarray}
and
\begin{eqnarray}
Q={\partial}_b-b{\partial}_t~,~\bar{Q}={\partial}_{\bar{b}}-\bar{b}{\partial}_{\bar{t}}.
\end{eqnarray}
In terms of $t$, $\bar{t}$ and $b$,$\bar{b}$ the $OSP(2,2)$ generators are given by 
\begin{eqnarray}
&&R_+=-{\partial}_t-{\bar{t}}^2{\partial}_{\bar{t}}-\bar{t}\bar{b}{\partial}_{\bar{b}}~,~R_-={\partial}_{\bar{t}}+{{t}}^2{\partial}_{{t}}+{t}{b}{\partial}_{{b}}~,~R_3=\bar{t}{\partial}_{\bar{t}}-t{\partial}_t+\frac{1}{2}\bar{b}{\partial}_{\bar{b}}-\frac{1}{2}b{\partial}_b\nonumber\\
&&V_+=\frac{1}{2}(Q+\bar{t}\bar{Q})~,~V_-=\frac{1}{2}(\bar{Q}-tQ),
\end{eqnarray}
and
\begin{eqnarray}
&&D_0=\bar{b}{\partial}_{\bar{b}}-b{\partial}_b~,~D_+=\frac{1}{2}(D-\bar{t}\bar{D})~,~D_-=-\frac{1}{2}(\bar{D}+tD).
\end{eqnarray}
Let us compute
\begin{eqnarray}
({\cal D}_0^2+2)A_+
&=&\frac{1}{2}\bar{b}\bar{\cal D}A+\frac{1}{2}b{\cal D}A+\bar{b}b\bar{\cal D}{\cal D}A-\bar{t}\bar{b}b\bar{\cal D}{\cal D}\bar{A}-\frac{1}{2}\bar{b}\bar{t}\bar{\cal D}\bar{A}-\frac{1}{2}b\bar{t}{\cal D}\bar{A}+A-\bar{t}\bar{A}.
\end{eqnarray}
\begin{eqnarray}
({\cal D}_0{\cal D}_++{\cal V}_+)(W)&=&-\frac{1}{2}\bar{b}b(\bar{t}{\cal D}\bar{\cal D}\bar{A}+\bar{\cal D}{\cal D}A)+\frac{3}{2}\bar{b}b({\cal D}^2\bar{A}+\bar{t}\bar{\cal D}^2A)+\frac{1}{2}(\bar{t}\bar{A}-A)+(\bar{t}b-\frac{1}{2}\bar{b})\bar{\cal D}A\nonumber\\
&+&(-\bar{b}+\frac{1}{2}\bar{t}b){\cal D}\bar{A}+\frac{1}{2}(b{\cal D}A-\bar{t}\bar{b}\bar{\cal D}\bar{A}).
\end{eqnarray}
In above we have used the identities ${\cal D}^2={\partial}_t$, $\bar{\cal D}^2={\partial}_{\bar{t}}$, ${\cal D}\bar{\cal D}={\partial}_b{\partial}_{\bar{b}}$, $\bar{\cal D}{\cal D}={\partial}_{\bar{b}}{\partial}_b$ and $\{{\cal D},\bar{\cal D}\}=0$. We can also compute
\begin{eqnarray}
-12{\cal D}_{-}({\cal D}_+A_+)&=&3\bar{\cal D}{\cal D}A_+-3\bar{t}t{\cal D}\bar{\cal D}A_++3t{\cal D}^2A_+-3\bar{t}\bar{\cal D}^2A_+-3\bar{b}\bar{\cal D}A_+\nonumber\\
&=&-\frac{3}{2}\bar{t}\bar{\cal D}{\cal D}\bar{A}-\frac{3}{2}\bar{t}t{\cal D}\bar{\cal D}A-\frac{3}{2}\bar{t}\bar{\cal D}^2A+\frac{3}{2}\bar{t}^2\bar{\cal D}^2\bar{A}+\frac{3}{2}t{\cal D}^2A+\frac{3}{2}\bar{\cal D}{\cal D}A-\frac{3}{2}t\bar{t}{\cal D}^2\bar{A}\nonumber\\
&-&\frac{3}{2}\bar{b}{\cal D}\bar{A}+\frac{3}{2}\bar{t}^2t{\cal D}\bar{\cal D}\bar{A}+\frac{3}{2}\bar{t}\bar{b}\bar{\cal D}\bar{A}+\frac{3}{2}\bar{t}\bar{A}-\frac{3}{2}\bar{b}\bar{\cal D}A-\frac{3}{2}\bar{t}t\bar{b}{\cal D}\bar{A}.
\end{eqnarray}
Also
\begin{eqnarray}
12{\cal D}_+^2A_- &=&3{\cal D}^2A_-+3\bar{t}^2\bar{\cal D}^2A_-+3\bar{t}\bar{b}\bar{\cal D}A_-\nonumber\\
&=&-\frac{3}{2}{\cal D}^2\bar{A}-\frac{3}{2}t{\cal D}^2A-\frac{3}{2}A-\frac{3}{2}\bar{t}^2\bar{\cal D}^2\bar{A}-\frac{3}{2}\bar{t}^2t\bar{\cal D}^2A-\frac{3}{2}\bar{t}\bar{b}\bar{\cal D}\bar{A}-\frac{3}{2}\bar{t}t\bar{b}\bar{\cal D}A.
\end{eqnarray}
We get immediately ( with $y=1+\bar{t}t+\bar{b}b$, $\omega=\bar{\cal D}A+{\cal D}\bar{A}$  and $2{\omega}_6=-\bar{t}b+\bar{b}$ )
\begin{eqnarray}
F_+&=&-\frac{3}{2}y\bigg({\cal D}^2\bar{A}+\bar{t}\bar{\cal D}^2A+\bar{t}\bar{\cal D}{\cal D}\bar{A}-\bar{\cal D}{\cal D}A+\bar{b}\omega\bigg)+(\bar{b}-\bar{t}b)\omega\nonumber\\
&=&-\frac{3}{2}\bigg({\cal D}(y\omega)+\bar{t}\bar{\cal D}(y\omega)\bigg)-4{\omega}_6\omega.
\end{eqnarray}
Since $F_+^{++}=F_-$ we must have ( by using also ${\cal D}^{++}=-\bar{\cal D}$, $\bar{\cal D}^{++}={\cal D}$, $\bar{A}^{++}=A$, $A^{++}=-\bar{A}$ and ${\omega}^{++}=\omega$, $2{\omega}_6^{++}=2{\omega}_7=-t\bar{b}-b$ )
\begin{eqnarray}
F_-=-\frac{3}{2}\bigg(\bar{\cal D}(y\omega)-{t}{\cal D}(y\omega)\bigg)-4{\omega}_7\omega.
\end{eqnarray}
Also 
\begin{eqnarray}
4{\cal D}_+{\cal D}_-W-4{\cal D}_-{\cal D}_+W&=&2(1+\bar{t}t)\bar{\cal D}{\cal D}W-b{\cal D}W-\bar{b}\bar{\cal D}W\nonumber\\
&=&2y(\bar{b}\bar{\cal D}{\cal D}\bar{A}-b\bar{\cal D}{\cal D}A)-2(1+\bar{t}t)\omega-W-\bar{b}b\omega.
\end{eqnarray}
\begin{eqnarray}
4{\cal V}_+A_--4{\cal V}_-A_+=2y(b{\cal D}^2\bar{A}+\bar{b}\bar{\cal D}^2A)-(1+\bar{t}t)\omega -W.
\end{eqnarray}
\begin{eqnarray}
-4{\cal D}_+{\cal D}_0A_-+4{\cal D}_-{\cal D}_0A_+&=&(1+\bar{t}t)({\cal D}{\cal D}_0\bar{A}-\bar{\cal D}{\cal D}_0A)+b{\cal D}_0A+\bar{b}{\cal D}_0\bar{A}\nonumber\\
&=&y\big(\bar{b}\bar{\cal D}{\cal D}\bar{A}-b\bar{\cal D}{\cal D}A+b{\cal D}^2\bar{A}+\bar{b}\bar{\cal D}^2A-\omega\big).
\end{eqnarray}
Hence
\begin{eqnarray}
f&=&3y(\bar{b}\bar{\cal D}{\cal D}\bar{A}-b\bar{\cal D}{\cal D}A+b{\cal D}^2\bar{A}+\bar{b}\bar{\cal D}^2A-2\omega)+2(y+\bar{b}b)\omega\nonumber\\
&=&3\big(b{\cal D}(y\omega)+\bar{b}\bar{\cal D}(y\omega)\big)-4(y+\bar{b}b)\omega.
\end{eqnarray}
We can immediately compute the Yang-Mills action
\begin{eqnarray}
S_{YM}[A]&=&\alpha STr(F_+F_--F_-F_++\frac{1}{4}f^2)\nonumber\\
&=&\alpha STr\bigg(\frac{9}{2}y{\cal D}(y\omega)\bar{\cal D}(y\omega)+4y^2{\omega}^2\bigg)\nonumber\\
&=&\frac{\alpha}{2\pi i}\int \frac{d\bar{t}dt d\bar{b}db}{y}\bigg(\frac{9}{2}y{\cal D}(y\omega)\bar{\cal D}(y\omega)+4y^2{\omega}^2\bigg).
\end{eqnarray}
In the last line we have also converted the supertrace into a superintegral. Similarly the Chern-Simons action becomes
\begin{eqnarray}
S_{CS}[A]&=&\beta STr(A_+F_--A_-F_++\frac{1}{4}Wf)\nonumber\\
&=&\beta STr\bigg(-\frac{3}{4}y\bigg(A\bar{\cal D}(y\omega)+\bar{A}{\cal D}(y\omega)\bigg)\bigg)\nonumber\\
&=&\frac{\beta}{2\pi i}\int \frac{d\bar{t}dt d\bar{b}db}{y}\bigg(-\frac{3}{4}y\bigg(A\bar{\cal D}(y\omega)+\bar{A}{\cal D}(y\omega)\bigg)\bigg).
\end{eqnarray}
The last step is to rewrite the above actions in terms of components of the superfields $A$ and $\bar{A}$. Introduce
\begin{eqnarray}
&&iA=\zeta +bv +\frac{1}{2}\bar{b}\frac{w+iu}{1+\bar{t}t}+\bar{b}b(\frac{\eta}{1+\bar{t}t}+{\partial}_t\bar{\zeta})\nonumber\\
&&i\bar{A}=-\bar{\zeta} +\bar{b}\bar{v} -\frac{1}{2}{b}\frac{w-iu}{1+\bar{t}t}+\bar{b}b(\frac{\bar{\eta}}{1+\bar{t}t}+{\partial}_{\bar{t}} {\zeta}).
\end{eqnarray}
$w$ and $u$ are real bosonic fields while $v$ is a complex bosonic field. Clearly $w^{++}=w,u^{++}=u, v^{++}=\bar{v}$. The fermionic fields $\zeta$ and $\eta$ are such that ${\zeta}^{++}=-\bar{\zeta}$, $\bar{\zeta}^{++}=\zeta$, ${\eta}^{++}=\bar{\eta}$, $\bar{\eta}^{++}=-\eta$. We can immediately compute
\begin{eqnarray}
iy\omega =iu+b\eta -\bar{b}\bar{\eta}+\bar{b}b\bigg((1+\bar{t}t)({\partial}_{\bar{t}}v-{\partial}_t\bar{v})+\frac{i}{1+\bar{t}t}u\bigg).
\end{eqnarray}
The Kahler term can now be put in the form ( with $\int d\bar{b}db \bar{b}b=-1$, $\int db=0$, $\int d\bar{b}=0$ )
\begin{eqnarray}
\frac{1}{2\pi i}\int \frac{d\bar{t}dt d\bar{b}db}{y}\bigg(\frac{9\alpha}{2}y{\cal D}(y\omega)\bar{\cal D}(y\omega)\bigg)&=&\frac{1}{2\pi i}\int d\bar{t}dt(\frac{9}{2}\alpha)\bigg(-(1+\bar{t}t)^2({\partial}_{\bar{t}}v-{\partial}_t\bar{v})^2+{\partial}_tu{\partial}_{\bar{t}}u\nonumber\\
&+&\frac{u^2}{(1+\bar{t}t)^2}+\eta {\partial}_{\bar{t}}\eta -{\partial}_t\bar{\eta}.\bar{\eta}-2iu({\partial}_{\bar{t}}v-{\partial}_t\bar{v})\bigg).
\end{eqnarray}
The superpotential takes the form
\begin{eqnarray}
\frac{1}{2\pi i}\int \frac{d\bar{t}dt d\bar{b}db}{y}(4\alpha y^2{\omega}^2)&=&\frac{1}{2\pi i}\int d\bar{t}dt(4\alpha)\bigg(\frac{2\bar{\eta}\eta}{1+\bar{t}t}-\frac{u^2}{(1+\bar{t}t)^2}+2iu({\partial}_{\bar{t}}v-{\partial}_t\bar{v})\bigg).
\end{eqnarray}
Simmilarly
\begin{eqnarray}
\frac{1}{2\pi i}\int \frac{d\bar{t}dt d\bar{b}db}{y}(-\frac{3}{4}\beta y)\bigg(A\bar{\cal D}(y\omega)+\bar{A}{\cal D}(y\omega)\bigg)&=&\frac{1}{2\pi i}\int d\bar{t}dt(-\frac{3}{4}\beta)\bigg(\frac{2\bar{\eta}\eta}{1+\bar{t}t}-\frac{u^2}{(1+\bar{t}t)^2}\nonumber\\
&+&2iu({\partial}_{\bar{t}}v-{\partial}_t\bar{v})\bigg).
\end{eqnarray}
In order to cancel the coupling between the fields $u$ and $v$ we choose $\beta=-\frac{2}{3}\alpha$. This will also cancel the mass term of the $u$ field. We obtain finally ( with $S_L[A]=S_{YM}[A]+S_{CS}[A]$ and $L{\longrightarrow}\infty $ )
\begin{eqnarray}
S_L[A]=\frac{1}{2\pi i}\int d\bar{t}dt(\frac{9}{2}\alpha)\bigg(-(1+\bar{t}t)^2({\partial}_{\bar{t}}v-{\partial}_t\bar{v})^2+{\partial}_tu{\partial}_{\bar{t}}u+\eta {\partial}_{\bar{t}}\eta +\bar{\eta}{\partial}_t\bar{\eta}+\frac{2\bar{\eta}\eta}{1+\bar{t}t}\bigg).
\end{eqnarray}

\section{Concluding remarks: A new fuzzy SUSY  scalar action}

The next natural step is to take the action (\ref{act2})  with the corresponding constraints (\ref{act3}) and (\ref{act4}) and write the whole thing in terms of the components of $X_{\pm,0},Y_i,Z_{\pm}$ thus  reducing the supertrace $STr$ to an ordinary trace $Tr$. The fermionic fields should   then be integrated out  before we can attempt any numerical investigation. This complicated exercise will not be pursuited here.

A possibly much simpler supersymmetric action than the above pure gauge action is given by the following fuzzy supersymmetric scalar action. We introduce the superscalar fields ${\Phi}_H$ and ${\Phi}_{\perp}$ defined by the expressions

\begin{eqnarray}
&&{\Phi}_H=(C_H+A_H)*\triangleleft (C_H+A_H)=-Y_i^2-Z_+Z_-+Z_-Z_+\nonumber\\
&&{\Phi}_{\perp}=(C_{\perp}+A_{\perp})*\triangleleft (C_{\perp}+A_{\perp})=\frac{1}{4}X_0^2+X_+X_--X_-X_+.
\end{eqnarray}
The action we write ( without any extra constraints and with full supersymmetry ) is
\begin{eqnarray}
S_L[{\Phi}]&=&STr(a_{\perp}{\Phi}_{\perp}+b_{\perp}{\Phi}_{\perp}^2+c_{\perp}{\Phi}_{\perp}^3+d_{\perp}{\Phi}_{\perp}^4+...)\nonumber\\
&+&STr(a_{H}{\Phi}_{H}+b_{H}{\Phi}_{H}^2+c_{H}{\Phi}_{H}^3+d_{H}{\Phi}_{H}^4+...).
\end{eqnarray}
This action is supersymmetric for the same reason that (\ref{act2}) is supersymmetric. It is gauge covariant since  ${\Phi}_H$ and ${\Phi}_{\perp}$ are gauge covariant fields. The parameters $a,b,c,d..$ are the coupling constants of the model. The partition function thus reads
\begin{eqnarray}
Z_L[a,b,c,d,...]=\int dX_{\pm}dX_0dZ_{\pm}dY_i e^{-S_L[\Phi]}.
\end{eqnarray}

We conclude this article by introducing  the matrix components of $X_{\pm,0},Y_i,Z_{\pm}$  in the following way. 
The superfields $X_0=D_0+W$ and $Y_i=R_i+C_i$ are real scalar superfields so that they are even elements of the superalgebra $Mat(2L+1,2L)$ while $X_{\pm}=D_{\pm}+A_{\pm}$ and $Z_{\pm}=V_{\pm}+B_{\pm}$ are odd elements of $Mat(2L+1,2L)$ ( we think of them as real spinor superfields ). This means that instead of considering the algebra $Mat(2L+1,2L)$ over the field of complex numbers ${\bf C}$ we consider it over a graded commutative algebra ${\bf P}$. Then the one-forms are actually elements of the space
\begin{eqnarray}
{\Psi}_N^1(P)=G_0\otimes Mat(2L+1,2L;P)_0 \oplus G_1\otimes Mat(2L+1,2L;P)_1.
\end{eqnarray} 
$G_0$ and $G_1$ are the even and odd parts of the super-Lie algebra $G=osp(2,2)$ while $Mat(2L+1,2L;P)_{0,1}$ are the subspaces of $Mat(2L+1,2L;P)$ with even and odd grading respectively with respect to the gradings of $Mat(2L+1,2L)$ and ${\bf P}$. ${\Psi}_N^1(P)$ is isomorphic to the space of one-forms ${\Psi}_N^1$ we had constructed previously.

Let us introduce the $(2L+1)\times (2L+1)$ fermionic matrices ${\psi}_{\pm R},{\chi}_{\pm R}$, the $2L\times 2L$ fermionic matrices ${\psi}_{\pm L}$, ${\chi}_{\pm L}$, the $(2L+1)\times 2L$ bosonic matrices $X_{\pm R},Z_{\pm R}$ and the $2L\times (2L+1)$ bosonic matrices $X_{\pm L},Z_{\pm L}$ as follows 
\begin{eqnarray}
&&X_{\pm}=\bigg(\begin{array}{cc}
                   {\psi}_{\pm R}& {X}_{\pm R} \\
            {X}_{\pm L} & {\psi}_{\pm L}  
                 \end{array}\bigg)~,~X_-={X}_+^{++}=\bigg(\begin{array}{cc}
                   {\psi}_{+R}^{++} & + {X}_{+L}^{++} \\
            - {X}_{+R}^{++} & {\psi}_{+L}^{++}
                 \end{array}\bigg).
\end{eqnarray}
 \begin{eqnarray}
&&Z_{\pm}=\bigg(\begin{array}{cc}
                   {\chi}_{\pm R}& {Z}_{\pm R} \\
            {Z}_{\pm L} & {\chi}_{\pm L}  
                 \end{array}\bigg)~,~Z_-={Z}_+^{++}=\bigg(\begin{array}{cc}
                   {\chi}_{+R}^{++} & + {Z}_{+L}^{++} \\
            - {Z}_{+R}^{++} & {\chi}_{+L}^{++}
                 \end{array}\bigg).
\end{eqnarray}
Furthermore we introduce the $(2L+1)\times (2L+1)$ bosonic matrices ${Y}_{iR}=Y_{iR}^{++},{X}_{0R}=X_{0R}^{++}$, the $2L\times 2L$ bosonic matrices ${Y}_{iL}=Y_{iL}^{++}$, ${X}_{0L}=X_{0L}^{++}$, the $(2L+1)\times 2L$ fermionic matrices ${\phi}_{iR}=-{\phi}_{iL}^{++},{\phi}_{0R}=-{\phi}_{0L}^{++}$ and the $2L\times (2L+1)$ fermionic matrices ${\phi}_{iL}={\phi}_{iR}^{++},{\phi}_{0L}={\phi}_{0R}^{++}$ as follows
\begin{eqnarray}
&&Y_i=\bigg(\begin{array}{cc}
                   {Y}_{iR}& {\phi}_{iR} \\
            {\phi}_{iL} & {Y}_{iL}  
                 \end{array}\bigg)~,~X_0=\bigg(\begin{array}{cc}
                   {X}_{0R}& {\phi}_{0R} \\
            {\phi}_{0L} & {X}_{0L}  
                 \end{array}\bigg).
\end{eqnarray}

 \paragraph{Acknowledgements}
The work of Badis Ydri is supported by a Marie Curie Fellowship
from The Commission of the European Communities ( The Research
Directorate-General ) under contract number MIF1-CT-2006-021797.

\bibliographystyle{unsrt}

\end{document}